\documentclass{article}

\usepackage{arxiv}
\usepackage[T1]{fontenc}
\usepackage{array,booktabs,multirow,tabularx}
\usepackage{amsmath,amssymb}
\usepackage{graphicx,adjustbox}
\usepackage{xcolor}
\usepackage{caption}
\usepackage{microtype}
\usepackage[numbers,sort&compress]{natbib}
\usepackage[colorlinks,urlcolor=blue,linkcolor=blue,citecolor=blue]{hyperref}

\newcolumntype{P}[1]{>{\raggedright\arraybackslash}p{#1}}
\newcolumntype{Y}{>{\raggedright\arraybackslash}X}
\renewcommand{\arraystretch}{1.08}

\renewcommand{\shorttitle}{LLM-Based Optimization for Wireless Communication}
\title{Large Language Model-based Optimization for Wireless Communication: A Survey}
\author{%
  Muhammad Sohaib\textsuperscript{1},
  Taehyung Kim\textsuperscript{1},
  Hyeri Heo\textsuperscript{2},\\
  Jemu Ryu\textsuperscript{2},
  Chaehyun Lee\textsuperscript{2},
  Jun Zhang\textsuperscript{1},
  and Sang-Woon Jeon\textsuperscript{1}\thanks{Corresponding author: Sang-Woon Jeon
    (\texttt{sangwoonjeon@hanyang.ac.kr}).}\\[0.6em]
  \textsuperscript{1}Department of Electrical and Electronic Engineering\\
  \textsuperscript{2}Department of Computer Science, College of Computing\\
  Hanyang University, Ansan 15588, South Korea\\[0.4em]
  {\small\normalfont\ttfamily
    msohaib828@hanyang.ac.kr, push0104@hanyang.ac.kr, junzhang@ieee.org}\\
  {\small\normalfont\ttfamily
    hwri12@hanyang.ac.kr, ryu1589@hanyang.ac.kr, dlcogus0919@hanyang.ac.kr}\\
  {\small\normalfont\ttfamily sangwoonjeon@hanyang.ac.kr}%
}
\date{}
\hypersetup{
  pdftitle={Large Language Model-based Optimization for Wireless Communication: A Survey},
  pdfkeywords={Large language model (LLM), wireless network optimization, resource management, prompt engineering, evolutionary computation, reinforcement learning.}
}

\begin{document}
\raggedbottom

\maketitle

\begin{abstract}
The growing complexity and dynamism of modern wireless communication frameworks have made real-time optimization essential for efficient resource allocation. Traditional optimization and machine-learning techniques often struggle with generalization, adaptability, stability, and scalability in heterogeneous, time-varying environments. Large language models (LLMs) have recently emerged as powerful tools capable of reasoning over structured prompts and generating context-aware decisions in real time without domain-specific retraining. This survey provides a comprehensive review of LLM-driven optimization techniques applied to wireless communication problems such as power control, bandwidth allocation, task offloading, and intelligent surface configuration. We introduce a functional taxonomy of LLM-enabled wireless resource management strategies and analyze the prompt engineering methodologies employed in recent literature. Furthermore, we develop a generalized analytical framework that categorizes LLM-based optimization architectures, including direct solution generation, closed-loop optimization with external verifiers, and representation encoding within trainable neural pipelines. To demonstrate practical applicability, we present representative case studies on LLM-assisted beam prediction for millimeter-wave systems and adaptive wireless resource allocation. Finally, we discuss emerging hybrid AI approaches for wireless networks.
\end{abstract}


\paragraph{Impact Statement}

This survey highlights the potential of large language models in enabling intelligent and autonomous wireless resource management for next-generation communication systems. By providing a unified taxonomy, analytical modeling framework, and representative case studies, this work establishes foundational guidelines for integrating language-driven intelligence into wireless optimization. The insights presented can support the development of AI-native and intent-driven network architectures, particularly for emerging 6G ecosystems. LLM-enabled optimization may improve spectral efficiency, scalability, and operational automation while reducing energy consumption and network management complexity. Furthermore, this work promotes interdisciplinary collaboration between wireless communications and artificial intelligence communities, accelerating the transition toward fully autonomous network management.

\section{Introduction}

The increasing demands for ultra-reliable and low-latency communication (URLLC), high data rates, and massive connectivity, which stem from emerging applications such as autonomous vehicles \cite{faran_integenaautoveh}, the massive Internet of Things (IoT) \cite{guo_enabmassiot}, augmented reality \cite{Morin_todisimaugme}, and the Tactile Internet \cite{hou_intelcomtactile}, have placed unprecedented pressure on the development of new technologies and the efficient utilization of resources in modern wireless networks. Optimization has emerged as a critical aspect in the design and management of wireless networks. A broad spectrum of techniques, ranging from convex optimization to heuristic algorithms and machine learning (ML), has been employed to tackle key challenges such as bandwidth allocation \cite{Xu_bandwidalloc}, power control \cite{wang_powercontrol}, task offloading in mobile edge computing (MEC) \cite{sohaib_hybridonoff}, scheduling \cite{zhang_devschedu}, and beamforming \cite{fredj_beamforming}. 
However, traditional optimization methods rely heavily on precise system models, scale poorly, and lack adaptability in dynamic environments. ML methods, while promising, face challenges such as high data requirements, limited interpretability, and difficulty generalizing across diverse network conditions.

Recent advances in large language models (LLMs) \cite{Hagos_recadvllmapp}, such as GPT and LLaMA, have opened new possibilities for tackling wireless optimization problems from a novel perspective. Originally designed for natural language understanding and generation, LLMs have demonstrated strong capabilities in complex reasoning, planning tasks, and contextual understanding. This has sparked a new line of research exploring the use of LLMs as general-purpose optimizers that operate on structured prompts and can produce feasible decisions without requiring domain-specific training \cite{Long_surveyllmoptim}. In wireless communications, this shift introduces a prompt-based optimization paradigm that is both flexible and interpretable. 
Although recent studies have highlighted the potential of LLMs in enhancing wireless networks, including their applications in optimization and management tasks \cite{zhou_LLMfTelecomsurvey,Javaid_leverLLMISATN,Tang_utiliLLMadvopt}, they primarily provide conceptual frameworks, dataset resources, or theoretical discussions. These works do not report actual implementation case studies where LLMs are directly applied to real-world optimization problems within wireless networks.

{\color{black}
This survey focuses on LLM-enabled wireless resource optimization rather than general LLM applications in telecommunications. We examine how LLMs contribute to wireless resource-management workflows through direct optimization, state/context modeling, objective and constraint interpretation, tool-augmented optimization, and automation/control-interface support. Representative tasks include power control, bandwidth allocation, MEC, beamforming, traffic prediction, BS/RIS/UAV deployment, ISAC, standards understanding, intent processing, and UAV-assisted networks. Compared with existing surveys that mainly discuss potential LLM applications, this work provides a focused optimization-oriented perspective by jointly considering task taxonomy, LLM pipeline roles, analytical modeling, prompting strategies, optimization objectives and constraints, evaluation mechanisms, metrics, and deployment timescales, as summarized in Table~\ref{tab:comparison_contribution}.

The main contributions of this survey are summarized as follows:
\begin{itemize}
    \item We provide a comprehensive overview of LLM-enabled wireless resource optimization, with emphasis on pipeline roles, practical case studies, quantitative evaluation, and deployment feasibility.
    
    \item We propose a functional and pipeline-role taxonomy covering power control, bandwidth allocation, mobile edge computing, beamforming, traffic prediction, network deployment, ISAC, standards understanding, intent processing, and UAV-assisted networks.
    
    \item We present a unified analytical modeling framework for LLM-based wireless optimization, covering direct optimization, state/context modeling support, objective/constraint interpretation support, tool-augmented optimization, closed-loop feedback, and trainable neural integration.
    
    \item We analyze prompting, context engineering, and harness engineering strategies and relate them to wireless optimization tasks, system requirements, external tools, simulators, and verifiers.
    
    \item We synthesize the surveyed works in terms of optimization objectives, constraints, validation mechanisms, metrics, deployment timescales, and hybrid AI integration with ML, DRL, evolutionary computation, solvers, and simulators.
\end{itemize}
}

The rest of the paper is organized as follows. Section II reviews the fundamentals of LLMs and wireless network optimization and introduces the motivation for using LLMs in this context. Section III presents a functional taxonomy of LLM-based wireless resource management approaches. Section IV analyzes prompt engineering methods used in existing studies. Section V introduces generalized analytical models for LLM-based wireless resource optimization, presents representative case studies (including beam prediction and adaptive resource allocation), and discusses techniques for real-time and energy-efficient LLM deployment. Section VI discusses hybrid approaches that integrate LLMs with DRL, ML, and evolutionary computation techniques. Finally, Section VII concludes the survey and outlines open challenges and future research directions.

\begin{table}[t]
\centering
\caption{Comparison of This Work with Existing Surveys}
\label{tab:comparison_contribution}
\renewcommand{\arraystretch}{1.02}
\begin{adjustbox}{max width=\linewidth}
\begin{tabular}{|c|c|c|c|c|c|c|c|}
\hline
\textbf{Refs.} & 
\multicolumn{7}{c|}{\textbf{Key Aspects of LLM-Driven Wireless Optimization}} \\ \cline{2-8}

& \textbf{Prompt} & \textbf{Analytical} & \textbf{Case} & \textbf{Wireless} & \textbf{LLM} & \textbf{Quantitative} & \textbf{Hybrid} \\
& \textbf{Engineering} & \textbf{Framework} & \textbf{Studies} & \textbf{Tasks} & \textbf{Architectures} & \textbf{Metrics} & \textbf{AI Integration} \\
\hline
{\cite{Long_surveyllmoptim}} &  &  &  & $\checkmark$ & $\checkmark$ &  &  \\
\hline
{\cite{zhou_LLMfTelecomsurvey}} &  &  &$\checkmark$  & $\checkmark$ & $\checkmark$ &  & $\checkmark$ \\
\hline
{\cite{Javaid_leverLLMISATN}} &  &  &  & $\checkmark$ & $\checkmark$ &  &  \\
\hline
{\cite{Boateng_ASLLMFCNASM}} &  &  &  & $\checkmark$ & $\checkmark$ &  & $\checkmark$ \\
\hline
{\cite{jiang_ACSOLAIMFFC}} &  &  & $\checkmark$  & $\checkmark$ & $\checkmark$ & $\checkmark$ &  \\
\hline
{\cite{Yang_DMLLMFWCAC}} & $\checkmark$ &  &  & $\checkmark$ & $\checkmark$ &  & $\checkmark$ \\
\hline
{\cite{wei2025largelanguagemodelsnextgeneration}} & $\checkmark$ &  & $\checkmark$ & $\checkmark$ & $\checkmark$ & $\checkmark$ &  \\
\hline
{\cite{zhou_LLMFWNAOF}} & $\checkmark$ &  & $\checkmark$ & $\checkmark$ &  & $\checkmark$  &  \\
\hline
{\cite{Javaid_LLMFUAV}} &  &  &  & $\checkmark$ & $\checkmark$ &  &  \\
\hline
{\cite{Nazar_NextGPTLGAFAWN}} &  &  & $\checkmark$  & $\checkmark$ & $\checkmark$ & $\checkmark$ & \\

\hline
\textbf{Our Work} & $\checkmark$ & $\checkmark$ & $\checkmark$ & $\checkmark$ & $\checkmark$ & $\checkmark$ & $\checkmark$ \\
\hline

\end{tabular}
\end{adjustbox}
\end{table}

\section{Background: Large Language Models and Wireless Network Optimization}
\subsection{Fundamentals of Large Language Models}
LLMs are deep neural networks (DNNs) which are trained on a large corpus of text using self-supervised ML to understand and generate human-like text. The capabilities of these models depend on the number of parameters, the size of the training data, and the computational resources utilized during training. Here, the parameters refer to adjustable weights within neural networks that are fine-tuned during training to enable the model to predict the next word in a sequence. Because language is inherently sequential, with meaning depending on word order and context, LLMs are designed to predict the next word based on preceding ones. LLMs are able to make these predictions by using an architecture called ``Transformer".
\begin{figure}[htbp]
    \centering
    \includegraphics[width=0.65\linewidth]{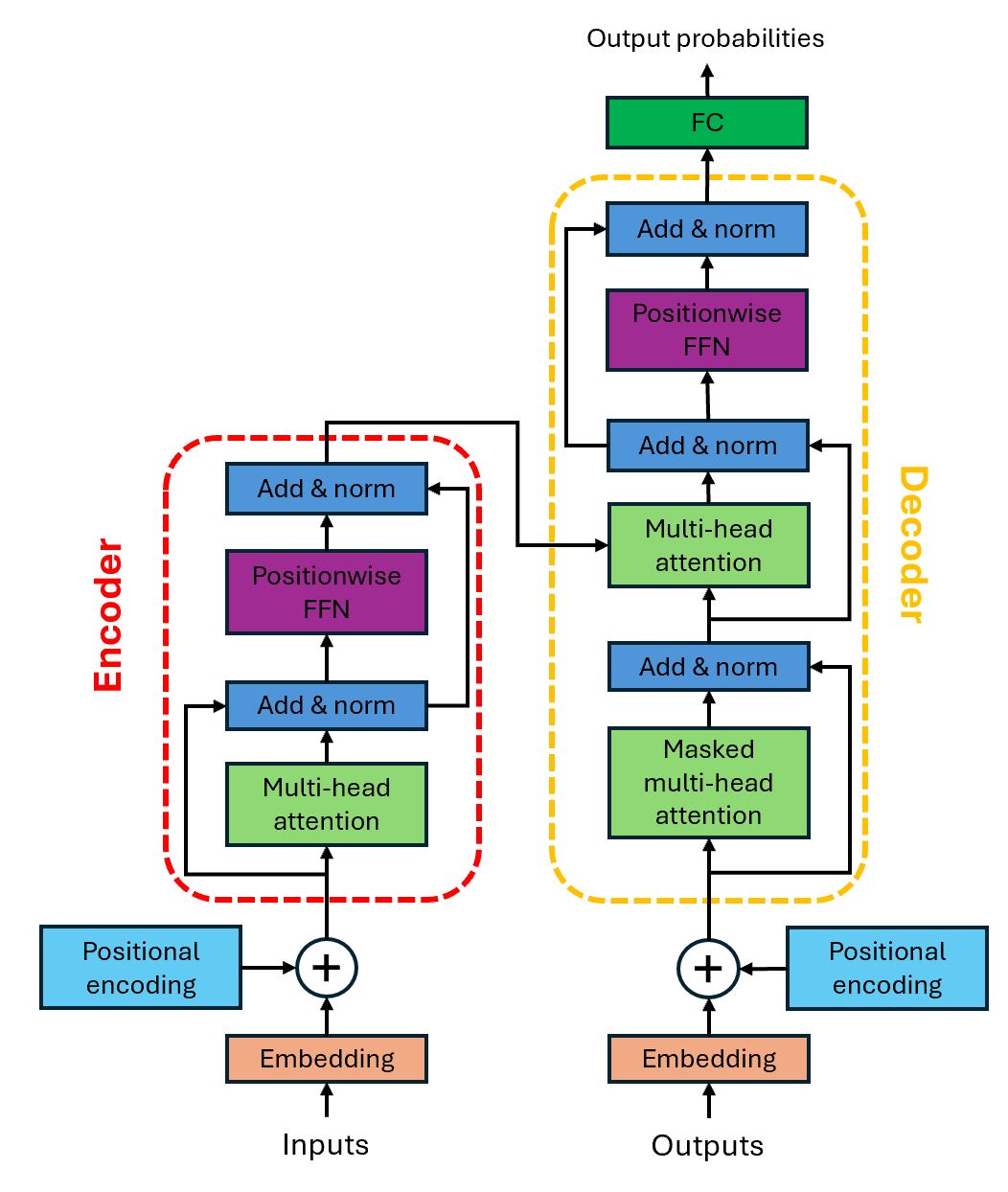}
    \caption{Transformer architecture.}
    \label{fig:transformer}
\end{figure}

The Transformer architecture, as shown in Fig. \ref{fig:transformer}, was developed for machine translation and comprises two main components: an encoder and a decoder \cite{vaswani2017attention}.
The encoder's role is to process the input text and capture contextual information by converting it into a sequence of numerical vector representations. The decoder then uses this vector sequence to generate the output text. In a translation task, the encoder converts source-language text into vectors, which the decoder then uses to generate text in the target language. Both the encoder and decoder consist of multiple layers connected through a self-attention mechanism, which allows the model to evaluate the importance of each word in a sequence relative to others. This mechanism enables the model to capture long-range dependencies and contextual relationships across the input, helping it generate output that is more coherent and contextually meaningful. Based on the encoder-decoder architecture, LLMs are generally classified into three categories: encoder-only, encoder-decoder, and decoder-only models \cite{zhou2024comprehensiveLLM}.

Encoder-only models, such as BERT, RoBERTa, DistilBERT, and ALBERT, focus on understanding input text by capturing deep contextual representations. They are primarily used for tasks like text classification, named entity recognition, and question answering. Encoder-decoder models, like T5, PEGASUS, MarianMT, and BART, are designed for sequence-to-sequence tasks where both understanding and generation are required such as text summarization, translation and paraphrasing. Decoder-only models, notably GPT  and LLaMA variants such as GPT-3, GPT-4, LLaMA-2, and LLaMA-3, generate text by predicting the next word in a sequence and are widely used in open-ended text generation, dialogue systems, and code completion.

Beyond their language generation capabilities, modern LLMs have demonstrated strong reasoning and decision-making abilities through mechanisms such as in-context learning and chain-of-thought (CoT) prompting. These models can process structured prompts containing task descriptions, constraints, and historical examples, and generate coherent outputs that approximate optimization solutions. Techniques like zero-shot and few-shot prompting allow LLMs to adapt to new tasks without retraining, making them attractive for dynamic and constraint-rich domains such as wireless communication systems.
Table \ref{tab:llm_applications} provides an overview of domain-specific LLMs across key sectors including finance, healthcare, and law, as well as education and industry~\cite{ref2-16}. 
Each model varies in size, training scale, and architecture, reflecting design choices that are tailored to domain-specific needs and constraints. These design choices include techniques such as fine-tuning, retrieval augmentation, and parameter-efficient adaptation~\cite{ref2-17,ref2-18}. This illustrates how LLMs are increasingly leveraged to deliver scalable and specialized solutions that exceed the capabilities of general-purpose optimizers, following the broader trend toward domain-centric intelligence~\cite{ref2-19}.

While domain-specific LLMs have matured in sectors such as healthcare and finance, telecom-oriented large models are only recently emerging. Several efforts have begun to develop foundation models tailored for wireless communication and network operations. For example, WirelessMathLM introduces domain-specific reinforcement learning to teach LLMs mathematical reasoning for wireless communication problems involving optimization constraints and signal-processing formulations \cite{li2025wirelessmathlm}. Similarly, RF-GPT proposes a radio-frequency language model that integrates RF spectrogram representations into multimodal LLMs to enable RF-aware reasoning and wireless signal understanding \cite{zou2026rf}. In addition, frameworks such as NetOrchLLM utilize LLMs as orchestration engines that coordinate multiple specialized wireless models for tasks such as network management and resource optimization \cite{abdallah2024netorchllm}. 

Beyond academic research, industry efforts are also exploring telecom-specific foundation models. For instance, SoftBank has recently introduced a Large Telecom Model (LTM) trained on operational network data to assist in cellular network design, management, and base station configuration optimization \cite{softbank2025ltm}. Likewise, NVIDIA has introduced telecom reasoning models based on Nemotron architectures to support autonomous network operations and agentic workflows for network management and optimization \cite{dendukuri2026nvidia}. These developments highlight a growing trend toward telecom-specialized large models that integrate domain knowledge, network telemetry, and wireless signal data into LLM-based reasoning frameworks.

\begin{table}[!t]
\centering
\small
\renewcommand{\arraystretch}{1.02}
\caption{Representative domain-specific LLMs across industries with architectural and adaptation characteristics.}
\label{tab:llm_applications}
\begin{adjustbox}{max width=\linewidth}
\begin{tabular}{|
p{2.7cm}|
p{2.7cm}|
p{1.6cm}|
p{2.5cm}|
p{5.6cm}|}
\hline
\textbf{Domain} & \textbf{Model} & \textbf{Parameters} & \textbf{Base Model} & \textbf{Adaptation / Notes} \\
\hline

\multirow{3}{*}{Healthcare}
& Med-PaLM 2 \cite{singhal_medpalm2} & 340 billion & PaLM 2 & Medical QA specialization with expert-level evaluation \\
& BioMistral \cite{labrak_biomistral} & 7 billion & Mistral 7B & Medical QA specialization with multilingual capability \\
& ChatDoctor \cite{li_chatdoctorllama} & 7 billion & LLaMA & Instruction-tuned medical dialogue system \\

\hline

\multirow{2}{*}{Finance}
& BloombergGPT \cite{wu2023bloomberggpt} & 50 billion & BLOOM & Financial QA and suggestion of news headlines \\
& FinGPT \cite{yang2025fingpt} & 8.3 million & LLaMA-3.1-8B-Instruct & Financial sentiment analysis, market understanding, and robo-advisory applications \\

\hline

\multirow{2}{*}{Legal}
& SaulLM \cite{colombo_saulLM} & 141 billion & Mistral-141B-Instruct & Legal reasoning, case analysis, and structured interpretation of legal documents \\
& LawLLM \cite{dong_lawLLM} & 7 billion & Gemma-7B & Legal case retrieval, precedent recommendation, and judgment prediction \\

\hline

\multirow{2}{*}{Education}
& EduChat \cite{dinh_educhat} & 175 billion & ChatGPT & Tutoring-oriented dialogue systems \\
& MathCoder \cite{wang_mathcoder} & 7--70 billion & LLaMA-2 & Math and code reasoning specialization \\

\hline

\multirow{2}{*}{Customer Service}
& EcomGPT \cite{Li_EcomGPT} & 7.1 billion & BLOOMZ & E-commerce semantic understanding, query processing, and product content generation \\
& CSDM \cite{Yun_csdm} & 5.8 billion & Polyglot-Ko & Customer service conversation summarization and key issue extraction \\

\hline

\multirow{3}{*}{Telecommunications}
& Nemotron Telco Model \cite{dendukuri2026nvidia} & 30 billion & Nvidia Nemotron & Telecom network reasoning, workflow automation, and autonomous decision-making for network operations \\
& RF-GPT \cite{zou2026rf} & 7 billion & Qwen2.5-VL & Radio-frequency signal understanding, reasoning, and multimodal QA over wireless spectrograms \\
& WirelessMathLM \cite{li2025wirelessmathlm} & 0.5--7 billion & Qwen2.5  & Wireless communication math solving, equation completion, and structured reasoning \\

\hline

\multirow{2}{*}{Agriculture}
& AgriGPT \cite{yang_agrigpt} & 8 billion & Qwen3-8B & Agricultural QA, expert advisory, and knowledge-grounded reasoning \\
& AgroLM \cite{awais_agrogpt} & 7 billion & LLaVA & Multimodal agricultural QA, disease identification, and advisory reasoning from images \\

\hline

\multirow{2}{*}{Industry}
& MiningGPT \cite{fernando_mininggpt} & 7 billion & Mistral-7B-Instruct & Mining-domain QA, commonsense reasoning, and operational knowledge assistance \\
& SemiKong \cite{nguyen_semikong} & 70 billion & LLaMA-3 & Semiconductor manufacturing QA, process optimization, and expert-level knowledge assistance\\

\hline

\end{tabular}
\end{adjustbox}
\end{table}

\subsection{Optimization for Wireless Networks}
Wireless communication has rapidly evolved to support emerging applications such as smart cities, autonomous transport, and AI-driven systems, all of which demand high data rates, ultra-low latency, energy efficiency, and massive connectivity. While 5G networks are now mainstream, the focus is shifting to 6G, which aims to deliver terabit-per-second speeds, sub-millisecond latencies, and ultra-dense connectivity. These goals are enabled by innovations like massive MIMO with beamforming \cite{Ning_beammimo} and symbol-level precoding \cite{Li_sympremimo}, advanced waveforms, novel access protocols, cloud and cell-free architectures, spectrum sensing, and emerging technologies like intelligent reflecting surfaces \cite{najam_rishbeam} and AI-driven network optimization.

Mathematical optimization, a powerful and indispensable tool for modeling and problem-solving, lies at the core of all the aforementioned wireless communication technologies \cite{Liu_suroptimeth}. The mathematical characteristics of these optimization problems vary significantly depending on the nature of the decision variables and constraint functions. For example, they can be categorized as linear vs nonlinear, unconstrained vs constrained, smooth vs nonsmooth, convex vs nonconvex, stochastic vs deterministic, integer vs continuous vs mixed-variable, and polynomial-time solvable vs NP-hard. Further distinctions include static vs dynamic, single-objective vs multi-objective, centralized vs distributed, and deterministic vs stochastic optimization.

The choice of optimization methods for these problems is largely influenced by the underlying problem structure. For example, in mmWave-NOMA with imperfect SIC, a non-convex framework integrates cross-entropy-based clustering with convex optimization to manage interference trade-offs \cite{Lim_JUCBPAmNISIC}. Robust polynomial beamformer design employs high-dimensional semidefinite programming solved via alternating direction method of multipliers (ADMM) for computational efficiency \cite{Xue_drpvadmm}, while adaptive beamforming under nonconvex uncertainty uses worst-case signal-to-interference-plus-noise ratio (SINR) maximization reformulated as a quadratic matrix inequality, addressed via linear matrix inequality (LMI) and bilinear matrix inequality (BLMI) techniques \cite{Huang_rabvwsinrmnuc}. Joint BS and intelligent reflecting surface deployment is cast as a mixed-integer combinatorial problem solved through graph-based heuristics and branch-and-bound \cite{Mei_jbsirsncgraph}. In wideband integrated sensing and communication (ISAC) systems, a mixed-integer nonlinear problem is solved using generalized Benders' decomposition to jointly optimize subcarrier and beamformer design \cite{Wang_oradwisac}. For URLLC traffic scheduling, a complex mixed-integer nonlinear program (MINLP) is reduced to a pairing problem and handled with efficient matching algorithms \cite{almekhlafi_sburrlctraffic}. These mathematical optimization methods rely heavily on system modeling, struggle to scale with network size, lack flexibility in dynamic environments, and offer limited adaptability to changing conditions.

In addition to these traditional optimization techniques, ML approaches, particularly DNNs and DRL, have emerged as effective data-driven tools for solving complex wireless communication problems. A hybrid learning framework combining offline DNN-based delay estimation with online DRL decision-making has been proposed for efficient task offloading in dynamic MEC environments \cite{sohaib_hybridonoff}. Multi-agent reinforcement learning was applied to optimize dynamic multichannel access, offering provable guarantees on both throughput and fairness in wireless networks \cite{sohaib_dmamarl}. In addition, DRL has been applied for the autonomous deployment of drone base stations to enhance wireless coverage and user connectivity in dynamic environments \cite{Tarekegn_drldrbasestation}. ML methods also come with their own limitations, including high data dependence, stability and convergence challenges, complex model management and monitoring requirements, limited interpretability, and difficulties in generalization across varying network conditions.

The growing complexity and dynamism of wireless networks, particularly in 6G contexts, have exposed limitations in conventional optimization pipelines. There is a growing need for more flexible, generalizable, and interpretable optimization frameworks that can handle heterogeneous constraints and operate without extensive retraining. This has led to the recent exploration of attention-based neural networks, i.e., LLMs, as reasoning engines capable of generating feasible optimization decisions directly from structured prompts. In the next section, we discuss how LLMs are being utilized for meeting the unique demands of wireless optimization tasks.

\section{LLM-Enabled Wireless Resource Management: A Functional Taxonomy}
\label{sec:LLM_taxonomy}
\begin{figure}[!t]
    \centering
    \includegraphics[width=\textwidth]{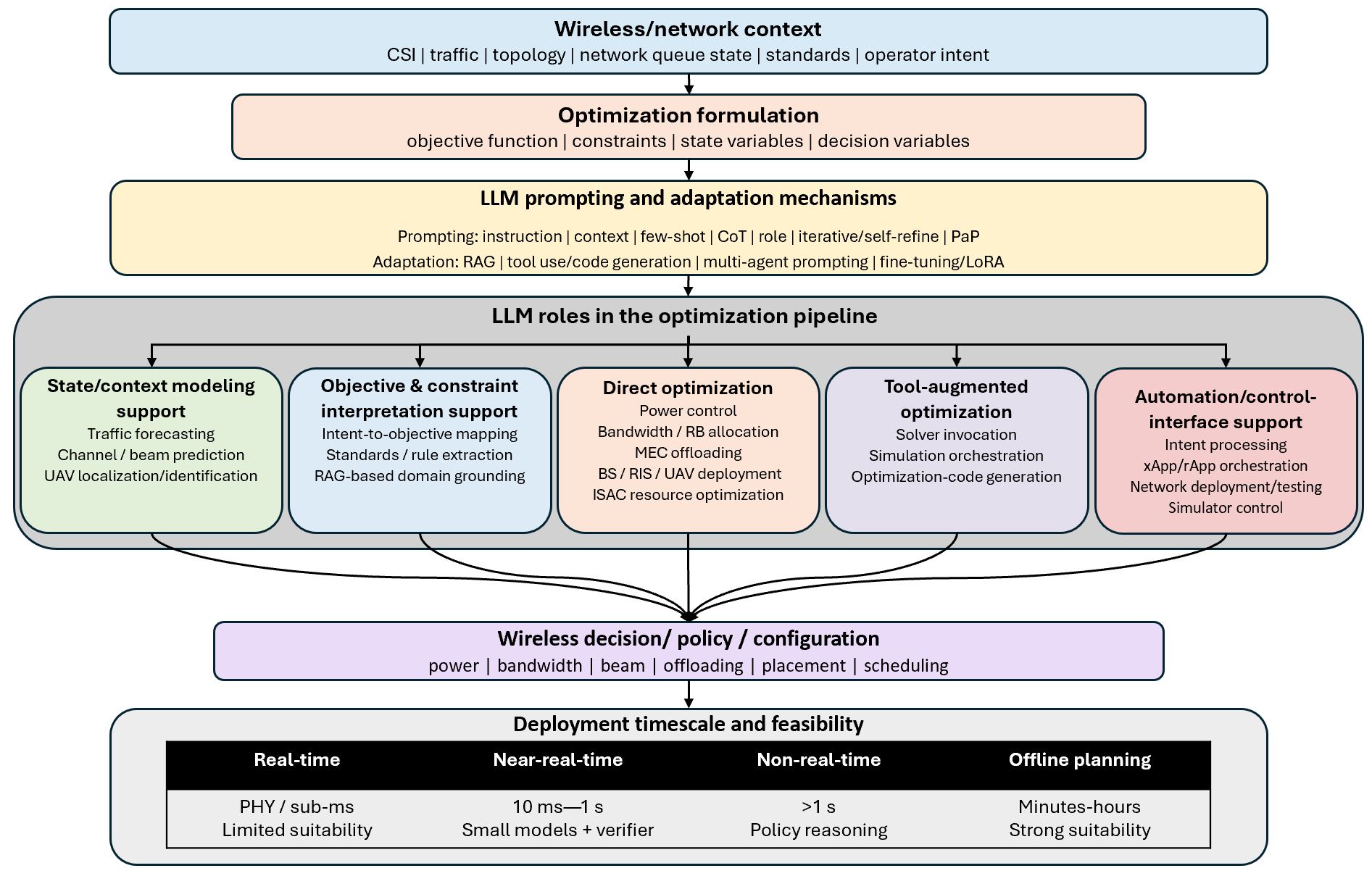}
    \caption{\textcolor{black}{Pipeline-level synthesis of LLM roles in wireless optimization. Wireless/network context is first abstracted into an optimization formulation defined by objectives, constraints, state variables, and decision variables. Prompting and adaptation mechanisms then enable different LLM roles, including state/context modeling support, objective and constraint interpretation support, direct optimization, tool-augmented optimization, and automation/control-interface support. These roles contribute to the generation, refinement, or execution of wireless decisions, policies, and configurations, which are finally mapped to deployment timescales ranging from real-time control to offline planning.}}
    \label{fig:llm_optimization_pipeline}
\end{figure}

{\color{black}
LLMs are increasingly being deployed for resource optimization in wireless communication systems because of their ability to understand natural-language instructions, generate and debug code, reason over mathematical and logical constraints, and interact with external tools or simulators. These capabilities allow LLMs to support diverse wireless optimization tasks, including power control, bandwidth allocation, base-station and UAV deployment, beam prediction, mobile edge computing, and integrated sensing and communication under dynamic network conditions. To clarify how these studies contribute to wireless optimization, Fig.~\ref{fig:llm_optimization_pipeline} presents a pipeline-level synthesis of LLM roles in the optimization workflow. The figure distinguishes the wireless/network context from the resulting optimization formulation, including the objective function, constraints, state variables, and decision variables. It further shows that LLMs can participate in the optimization pipeline through state/context modeling support, objective and constraint interpretation support, direct optimization, tool-augmented optimization, and automation/control-interface support. Building on this synthesis, Table~\ref{tab:llm_functional_summary} provides a functional taxonomy of representative studies according to both the wireless optimization domain and the pipeline role of the LLM. Each subsection then discusses the practical motivation, representative case studies, and key prompting or adaptation strategies used in each domain.
}

\subsection{Power Control}
Transmit power optimization is extremely important in wireless networks because of its effects on co-channel interference, signal-to-noise ratio (SNR) which affects quality of service (QoS), and sustainable communication because of IoT and mobile devices being run on batteries. Classical methods such as convex optimization and ML require problem-specific formulation, transformation of non-convex problems, laborious model training, and hyperparameter fine-tuning. In contrast, LLMs can overcome these limitations by using their zero-training inference abilities, language-based programmability and in-context learning.

In~\cite{zhou2024powercontrol}, an in-context learning algorithm was introduced for BS power control. By prompting the LLM with task goals and rule-based descriptions, it achieved near-optimal performance in both discrete and continuous decision environments without model retraining. The LLM adapts dynamically to network states using an experience pool.
Building on this, ~\cite{lee2024llmresource} designed a few-shot prompting strategy for dual-transmitter systems, focusing on energy and spectral efficiency. Their work evaluated various LLaMA models, showing that fine-tuned code-oriented LLMs yield better performance in transmit power prediction. A binary fallback rule was integrated to enhance robustness.
Using a combination of CoT prompting and program-aided language (PAL) models,~\cite{shao2024wirelessllm} generated executable MATLAB code for the water-filling algorithm for power control in an orthogonal frequency-division multiplexing (OFDM) system, demonstrating LLM reasoning across algorithmic and physical-layer (PHY-layer) procedures.
The study in~\cite{zou2025genainet} explored power control from a distributed multi-agent lens. Here, LLM agents coordinate power allocation by using prompts to share information and make decisions collaboratively, enabling efficient cooperative intelligence.
An LLM-gated mixture-of-experts (MoE) framework was proposed in~\cite{du2024moe}, where the LLM dynamically selects DRL experts to balance power-utility tradeoffs based on scenario-specific constraints.

In open radio access networks (O-RAN), LLMs have also been utilized to enhance intelligent control, enabling adaptive and data-driven decision making across non-real-time (non-RT) and near-real-time (near-RT) RAN layers. In ~\cite{bao2025llmhric}, LLMs were integrated into non-RT RAN intelligent controllers (RICs) to provide strategic guidance, using instructional prompting, to RL agents in near-RT RICs for power allocation in order to achieve higher total throughput.
~\cite{liu2025llmris} leveraged an LLM to jointly optimize the resource block (RB) allocation and decoding order of vehicles in an internet of vehicles (IoV) network. Based on the RB allocation and decoding order obtained from the LLM, they optimized RIS phase shifts and power in order to increase the sum-rate.
Lastly, an iterative prompting and self-refinement scheme was presented in~\cite{zhou2025promptengineering}, showcasing LLMs solving base station power minimization with constraints. The iterative framework improves LLM decisions using prior samples, achieving performance comparable to DRL in fewer steps.

\subsection{Bandwidth Allocation}
Due to the limited availability of radio frequency spectrum for wireless communications, efficient utilization of allocated network bandwidth is extremely important. Additionally, the explosive increase in the number of connected devices, high data rate demands, energy efficiency requirements, and high cost of expanding network infrastructure, have all played their part in making the bandwidth optimization even more critical. Traditional methods lack real-time adaptability; therefore recent studies have shown LLMs as flexible controllers for bandwidth assignment under traffic variation.

In~\cite{noh2025adaptive}, a closed-loop prompting framework called LLM-RAO was proposed for uplink resource allocation in IEEE 802.11ax. The LLM generates candidate allocations based on SINR, queue status, and fairness goals, refining them iteratively using external feedback.
In reconfigurable intelligent surface (RIS)-assisted vehicular networks, \cite{liu2025llmris} employed an LLM to assign RBs under mobility and CSI constraints. The model jointly optimized decoding order and RIS coefficients, enhancing sum-rate while satisfying reliability bounds.
A novel LLM-driven agentic AI framework was proposed in \cite{wu2025oranagent} for physical resource blocks (PRBs) allocation in O-RAN, with the objective of increasing the network reliability and utility. Meta-prompts, containing task description, historical optimization data, and contextual information, are iteratively given as an input to the LLM in order to generate optimal PRB allocations.

\subsection{Mobile Edge Computing}
Mobile edge computing has emerged as a powerful tool for handling computation-intensive and low-latency applications, such as autonomous systems, augmented reality, and scientific simulations. It brings computation and data storage systems closer to the resource-constrained devices (e.g., IoT, smartphones, vehicles). Recent studies have started utilizing LLMs for task allocation and user associations in MEC networks owing to their strong contextual  understanding and reasoning capabilities.

In~\cite{zhou2025edgecloud}, an edge–cloud collaboration strategy was proposed, where small LLMs (e.g., LLaMA-3-8B) serve delay-sensitive tasks at the network edge, while large models (e.g., GPT-4) process quality-critical tasks in the cloud. An in-context learning framework guides task offloading decisions using formatted examples and performance-based feedback. The LLM iteratively updates its meta-prompt with historical offloading outcomes and adapts decisions using prioritized experience replay and exploration strategies, converging to near-optimal delay–quality tradeoffs.
Similarly, \cite{song2024llmmec} tackled user-server assignment in MEC to minimize the maximum latency across MEC servers. Using GPT-4o-mini, the LLM receives natural language prompts encoding link processing times and task constraints, and infers allocation matrices. The prompts are refined over iterations using prior low-latency configurations. Compared to traditional model-based and heuristic baselines, the LLM quickly converges to near-optimal user–server assignments and generalizes across scenarios without retraining.

\subsection{Traffic Prediction and Analysis}
In order to maintain a consistently high QoS, traffic prediction and analysis are essential for proactive resource management because of growing dynamic and unpredictable network traffic. Traditional methods, such as time-series (e.g., RNN, LSTM) and graph-based learning, are unable to generalize and struggle with unpredictable traffic patterns. In contrast, LLMs offer enhanced capabilities owing to their ability to analyze diverse data types, such as network logs, usage statistics, through contextual prompts and instruction tuning.

In~\cite{shokouhi2024multitimespan}, a multi-timespan LLM framework was proposed that encodes spatial–temporal traffic histories into natural language prompts. Fine-tuned on BART, BigBird, and PEGASUS, the model achieves over $12\%$ improvement in mean absolute error (MAE) and up to $47\%$ better zero-shot generalization than graph neural network baselines. The prompts leverage both intra-cell history and auxiliary signals from clustered neighboring cells, offering robust multi-hour prediction under shifting traffic distributions.
A lightweight technique called self-refined prompting was introduced in~\cite{zhou2025promptengineering} for traffic prediction. By iteratively improving prior outputs, the LLM adapts its forecasting behavior without retraining. This method enables resource-efficient inference on constrained edge devices and aligns well with real-time prediction demands in wireless networks.

LLM applications in Wi-Fi environments were examined in~\cite{wang2024wifi}. A retrieval-augmented LLM assists in scenario-driven traffic assessment and parameter selection under high-density deployments. Paired with a generative diffusion model (GDM), the system performs traffic-aware channel and contention optimization in next-gen Wi-Fi networks.
Lastly, ~\cite{kan2024mobilellama} developed Mobile-LLaMA, a fine-tuned LLaMA-2 model for Network Data Analytics Function (NWDAF)-based traffic analytics. It supports IP routing and traffic monitoring in 5G core networks, generating scripts to analyze congestion, packet anomalies, and flow-level statistics. The model's domain-specific instruction tuning enables accurate traffic evaluation and capacity planning across user and slice contexts.

\subsection{Beam Prediction}
Beamforming, a core technology in MIMO, is a technique used to steer transmitted and received signals toward a specific user or device, rather than broadcasting in all directions. For non-stationary devices, traditional deep learning methods such as recurrent neural network (RNN)-based methods like LSTM, are unable to generalize and often require retraining for different network scenarios in order to provide a reliable high QoS.

To address this, ~\cite{sheng2025beam} reformulated beam prediction as a time-series forecasting task and proposed an LLM-based solution using the GPT-2 architecture. Their approach encodes historical optimal beam indices and angle of departure (AoD) into tokenized sequences via a patch embedding mechanism. A novel patch reprogramming technique then transforms these numerical features into discrete prototype tokens aligned with the vocabulary of the language model.
To enhance contextual awareness, the method incorporates prompt-as-prefix (PaP) prompts, which encode domain knowledge, statistical trends (e.g., signal drift), and lag structures. These prompts enable the LLM to reason over time-evolving beam dynamics, providing future beam indices with high accuracy and robustness.
Compared to LSTM-based models, the proposed method demonstrates improved robustness under mismatched antenna configurations, BS locations, and center frequencies. Notably, it achieves strong generalization without fine-tuning when applied to previously unseen BSs and propagation conditions.

\subsection{Base Station and RIS Deployment}
The deployment of infrastructure elements such as base stations, reconfigurable intelligent surfaces (RISs), and unmanned aerial vehicles (UAVs) is critical to maximizing coverage, minimizing interference, and ensuring cost efficiency in modern wireless networks. Recent approaches have leveraged LLMs for flexible, real-time decision-making based on structured natural language prompts.

An LLM-driven framework for BS siting was proposed in~\cite{wang2024bssllm}, integrating prompt engineering and retrieval-augmented generation (RAG) to solve the BSS problem with minimal human intervention. The framework supports three strategies: Prompt-optimized LLM (PoL), an agent-based strategy (LaBa), and a collaborative multi-agent system (CLaBa). The agents iteratively refine placement decisions based on objectives described in natural language, enabling dynamic reconfiguration and reduced manual effort. Evaluations show improved accuracy and deployment efficiency compared to conventional heuristics.
In~\cite{sevim2024llmbs}, an LLM-assisted RL framework was introduced to guide BS deployment in urban scenarios. Using Deep Deterministic Policy Gradient (DDPG), the system fine-tunes a DistilBERT model to optimize BS locations and orientations based on ray-traced signal maps. The LLM improves environmental adaptability and reduces convergence time compared to CNN-only actors, demonstrating practical benefits in simulation-based urban planning.
For aerial RIS or BS deployment, ~\cite{wang2025uavllm} developed a novel prompting-based algorithm for multi-UAV placement in Integrated Access and Backhaul (IAB) networks. The LLM receives structured prompts encoding past UAV locations, utility scores, and constraints, and iteratively refines UAV positions. A temperature-decaying mechanism balances exploration and exploitation during prompt-based optimization. Results show that the LLM achieves near-optimal placements with fewer iterations than DRL, making it suitable for real-time deployments in dynamic environments.

\subsection{Integrated Sensing and Communication}
Integrated Sensing and Communication (ISAC) has emerged as a key technology in  wireless communication, due to precise and high-resolution sensing abilities in 6G networks because of massive antenna arrays and terahertz (THz) waves, enabling the unification of communication and physical world sensing over the same spectral resources. Traditional ISAC optimization frameworks often use complex multi-objective evolutionary algorithms (MOEAs) to balance sensing accuracy and communication utility, which usually suffer from high computational cost and slow convergence.

To overcome these limitations, ~\cite{li2025isacllm} proposed an LLM-enabled decomposition-based MOEA (LEDMA) for multi-UAV ISAC systems. The system aims to jointly optimize UAV placement and power allocation to maximize network utility and minimize localization error (Cram\'{e}r--Rao bounds). By integrating GPT-3.5 Turbo as a black-box search operator into the MOEA/D framework, structured prompts are designed to encode solution history, fitness scores, and optimization constraints. These prompts are used iteratively to refine UAV deployment strategies without requiring retraining.
The LLM generates new solutions within constraints using in-context learning with natural language prompts. Compared to standard MOEAs such as NSGA-II and RVEA, LEDMA achieves superior convergence and Pareto front diversity. This demonstrates the potential of prompt-driven LLMs to assist in real-time ISAC system optimization, especially when balancing trade-offs in resource-constrained UAV deployments.

\subsection{Low-Altitude Economy and UAV-Assisted Wireless Networks}
The low-altitude economy (LAE) has received significant interest from both academia and industry for next-generation wireless networks. It combines terrestrial, non-terrestrial, and air-ground systems to provide a wide variety of services ranging from environmental monitoring and urban air mobility to logistics and smart-city services. These applications require reliable wireless connectivity, intelligent resource management, and real-time coordination between aerial and terrestrial nodes \cite{LLM_LAE_xuzhuo}.

Recent studies have begun exploring the integration of large language models (LLMs) to enhance communication and network intelligence in LAE systems. Due to their strong reasoning and contextual understanding capabilities, LLMs can assist in complex wireless optimization tasks such as integrated sensing and communication (ISAC), UAV resource allocation, and network coordination in dynamic aerial environments \cite{llm_isac_lae_yang}. For example, LLM-based approaches have been proposed to optimize near-field communications in extremely large-scale MIMO systems, enabling efficient precoding and power allocation for UAV communications \cite{LLM_LAE_xuzhuo}.

LLMs have also been applied to several UAV network management tasks. These include secure communication optimization through intent prediction and reinforcement learning, dynamic communication scheduling in multi-UAV coordination systems, and intelligent data collection scheduling in UAV-assisted sensor networks. In addition, LLM-based frameworks have been investigated for UAV identification and security monitoring using radio-frequency fingerprinting techniques in integrated sensing and communication networks \cite{uavllmisac_zheng}.

Overall, these works demonstrate the potential of LLMs to support intelligent decision-making and optimization in UAV-assisted wireless systems. As LAE infrastructures continue to expand, LLM-driven approaches are expected to play an important role in enabling scalable and adaptive aerial communication networks.

\subsection{Communication Standards Understanding}
The rapid increase in the demand for wireless communication services has driven the development of highly complex and intelligent network architectures. To ensure seamless operation of these new technologies and global interoperability, comprehensive  standards, such as those produced by Third Generation Partnership Project (3GPP), have been developed. However, these standards are highly technical, modular, contain conditional logic, context-dependent terms, cross-referencing and dependencies. To address these issues, LLMs combined with RAG have emerged as a promising solution. In the following, we highlight some representative studies that leverage this approach.

In~\cite{alabbasi2025teleoracle}, a lightweight RAG system called TeleOracle was proposed, built on the Phi-2 small language model (SLM), and featuring semantic segmentation, hybrid retrieval, and long-context inference via SelfExtend. A semantic router directs queries to relevant telecom standard repositories, reducing hallucination and achieving $81.2\%$ accuracy, surpassing GPT-4o on telecom-specific benchmarks.
~\cite{nikbakht2024tspecllm} proposed TSpec-LLM, a comprehensive dataset comprising over 30,000 3GPP documents, amounting to 13.5 GB and approximately 535 million words, while preserving the technical structure using Markdown formatting. 
A simple RAG framework was designed with two main modules to support real-time interpretation of the dataset. When applied to the TSpec-LLM dataset, this RAG-based approach enhanced accuracy from $45\%$ to $71\%$.

From an optimization perspective, communication standards understanding plays a critical role in defining the feasible solution space and operational constraints of wireless systems. Many optimization problems, such as resource allocation, scheduling, and network configuration, are inherently constrained by protocol specifications (e.g., transmit power limits, scheduling formats, and signaling procedures defined in 3GPP standards). By enabling automated parsing and interpretation of such standards, LLM-based frameworks can translate textual specifications into machine-interpretable constraints and policies, thereby facilitating constraint-aware optimization and reducing the need for manual rule encoding. Consequently, standards understanding serves as an enabling component within LLM-driven optimization pipelines rather than an isolated task.

\subsection{Intent Processing and Network Automation}
Intent processing and network automation are building blocks of an intelligent and autonomous modern wireless network. Intent processing refers to the ability of the network to understand the operators' high-level instructions expressed in natural language to optimize resources such as power and bandwidth. Recent efforts have introduced systems that integrate LLMs as natural language interfaces, enabling automatic configuration, optimization, and troubleshooting of complex infrastructure in Open RAN, cloud-native 5G, and next-generation wireless systems.

For example, in~\cite{maxenti2025autorAN}, AutoRAN was introduced as a zero-touch deployment platform where LLMs convert operator intents into code-based templates for automated orchestration. Similarly, ~\cite{habib2025intentRAN} introduced HDTGA, a hierarchical decision transformer guided by LLM-extracted performance intents to manage and schedule network applications such as power optimization and traffic steering. In addition to that, in ~\cite{wu2025oranagent}, a framework called LLM-agentic AI xApp was presented, which leverages LLM agents as intelligent controllers within the O-RAN architecture to enable dynamic resource allocation across network slices. Experimental results on the OpenAI Cellular (OAIC) testbed demonstrate notable improvements in average data rates, slice reliability, and system responsiveness.
In the simulation domain, advanced simulators like Sionna require significant domain expertise to operate effectively. To address this, \cite{liu2024netllm} proposed a network-oriented LLM by leveraging parameter-efficient fine-tuning (PEFT) and RAG to enable natural language control of complex tasks such as ray-tracing, radio map generation, and channel impulse response computation, thereby enhancing usability and efficiency.

{\color{black}
The above taxonomy shows that LLMs contribute to wireless optimization in different ways, ranging from direct decision generation to prediction, standards interpretation, tool use, and network automation. Table~\ref{tab:optimization_oriented_summary} summarizes these roles from an optimization-oriented perspective by comparing representative objectives, constraints, evaluation methods, metrics, and deployment timescales.
}

\renewcommand{\arraystretch}{1.2}

\begin{table}[htbp]
\centering
\caption{A comprehensive overview of LLM techniques for key problems in wireless networks.}
\label{tab:llm_functional_summary}
\renewcommand{\arraystretch}{1.02}
\small
\resizebox{\textwidth}{!}{
\begin{tabular}{|P{2.4cm}|P{1.4cm}|P{2.5cm}|P{2.8cm}|P{2.7cm}|P{3.2cm}|P{4.2cm}|}
\hline
\textbf{Functional area} & \textbf{Key papers} &\textbf{\textcolor{black}{Pipeline role}} & \textbf{Prompting style} & \textbf{LLM role} & \textbf{LLM models used} & \textbf{Baseline methods compared} \\
\hline
\textbf{Power control} & \cite{zhou2024powercontrol} & \textcolor{black}{Direct optimizer} & Few-shot prompting & Prompt-driven power controller & LLaMA-3-8B-Instruct, GPT-3.5 Turbo, LLaMA-3-70B & DRL \\
\cline{2-7}
& \cite{zou2025genainet} & \textcolor{black}{Objective/constraint interpretation support}  & Instructional + contextual prompting & Prompt-driven power optimizer & GPT-3.5 & DPC, GenAI agents alone \\
\cline{2-7}
& \cite{shao2024wirelessllm} & \textcolor{black}{Tool augmented optimization}  & Instructional + Chain-of-thought + Contextual prompting & Tool-using code generator
 & GPT-4, GPT-3.5 & Optimal water-filling solution, full knowledge scheme \\
\cline{2-7}
& \cite{du2024moe} & \textcolor{black}{Objective/constraint interpretation support}  & Chain-of-Thought + Role-based prompting & Intent-processor and expert-selector & GPT-3.5 Turbo & Traditional gate network-enabled MoE approach \\
\cline{2-7}
& \cite{bao2025llmhric} & \textcolor{black}{Objective/constraint interpretation support}  & Instructional + role-based prompting & Prompt-based policy generator & LLaMA-3.1-8B-Instruct & DDPG with cosine decay noise, DDPG with linear decay noise, Equal power allocation \\
\cline{2-7}
& \cite{liu2025llmris} & \textcolor{black}{Objective/constraint interpretation support}  & Instructional + Contextual prompting & Prompt-guided RIS
controller & General-purpose LLMs & Random RIS phase configuration, Non-RIS assisted IoV, RIS with fixed strategies \\
\cline{2-7}
& \cite{zhou2025promptengineering} & \textcolor{black}{Direct optimizer}  & Instructional + few-shot + iterative prompting & Prompt-driven power optimizer  & LLaMA-3-7B, LLaMA-70B, GPT-4 & DRL-Baseline \\
\cline{2-7}
& \cite{lee2024llmresource} & \textcolor{black}{Direct optimizer}  & Few-shot + contextual prompting & Prompt-driven power predictor & CodeLLaMA-7B, LLaMA-2-7B, LLaMA-2-7B-32K-Instruct & Exhaustive search, Binary power control, Random allocation \\
\hline
\textbf{Bandwidth allocation} & \cite{noh2025adaptive} & \textcolor{black}{Direct optimizer}  & Contextual + instructional prompting + tool-using & Prompt-driven optimization agent & GPT-3.5 Turbo & DRL, Genetic, Round-robin \\
\cline{2-7}
& \cite{liu2025llmris} & \textcolor{black}{Objective/constraint interpretation support} & Instructional + contextual prompting & Prompt-guided RIS controller & General-purpose LLMs & Random RIS phase configuration, Non-RIS assisted IoV, RIS with fixed strategies \\
\cline{2-7}
& \cite{wu2025oranagent} & \textcolor{black}{Direct optimizer} & Instructional + contextual + iterative prompting & Prompt-driven resource allocator & GPT-4 & Random allocation, Equal provisioning, proportional resource allocation \\
\hline
\textbf{Mobile edge computing} & \cite{zhou2025edgecloud} & \textcolor{black}{Direct optimizer} & Few-shot + instructional prompting & Prompt-driven task offloader & LLaMA-3-8B, GPT-4o, Gemma 7B, Gemini 1.5 Pro & DRL-based offloading, DRL without exploration, DRL with prioritized experience replay \\
\cline{2-7}
& \cite{song2024llmmec} & \textcolor{black}{Direct optimizer}  & Instructional + iterative prompting & Prompt-driven user-BS associator & GPT-4o-mini & Genetic algorithm (GA), exhaustive optimal search \\
\hline
\textbf{Traffic prediction and analysis} & \cite{zhou2025promptengineering} & \textcolor{black}{State/context modeling support}  & Self-refined + contextual prompting & Self-correcting predictor &  GPT-4 & Ground truth, LSTM, ARIMA \\
\cline{2-7}
& \cite{shokouhi2024multitimespan} & \textcolor{black}{State/context modeling support} & Contextual + instructional prompting & Fine-tuned traffic predictor & BART, BigBird, PEGASUS & Adaptive graph convolutional recurrent network, ground truth \\
\cline{2-7}
& \cite{wang2024wifi} & \textcolor{black}{Objective/constraint interpretation support} & Instructional + contextual + iterative + RAG prompting & Prompt-driven optimizer utilizing RAG & General-purpose LLMs & IEEE 802.11, DDPG \\
\cline{2-7}
& \cite{kan2024mobilellama} & \textcolor{black}{State/context modeling support} & Instructional + Contextual prompting & Instruction-tuned Code Generator & LLaMA-2-13B + fine-tuned using LoRA & GPT-3.5 Turbo, Base LLaMA-2-7B,70B \\
\hline
\textbf{ Beamforming} & \cite{sheng2025beam} & \textcolor{black}{State/context modeling support} & Prompt-as-prefix + instructional prompting & Prefix-conditioned beam predictor & GPT-2 & LSTM-based ODE, Cascaded LSTM methods \\
\hline
\end{tabular}}
\end{table}

\begin{table}[htbp]
\centering
\ContinuedFloat
\caption{A comprehensive overview of LLM techniques for key problems in wireless networks (continued).}
\small
\resizebox{\textwidth}{!}{
\begin{tabular}{|P{2.8cm}|P{1.4cm}|P{2.5cm}|P{2.8cm}|P{2.7cm}|P{3.2cm}|P{4.2cm}|}
\hline
\textbf{Functional area} & \textbf{Key papers} & \textbf{\textcolor{black}{Pipeline role}} & \textbf{Prompt style} & \textbf{LLM role} & 
 \textbf{LLM models used} &\textbf{Baseline compared} \\
\hline

\multirow{3}{*}{\shortstack[l]{\textbf{Base station and} \\ \textbf{RIS deployment}}} 
& \cite{wang2024bssllm}   & \textcolor{black}{Direct optimizer}   & Instructional + iterative + RAG prompting & Prompt-based BSS optimizer, Autonomous BSS agent, Cooperative multiple BSS agents & LLaMA, Bloom, GPT  & Particle swarm optimization (PSO), simulated annealing (SA) \\
\cline{2-7}
& \cite{sevim2024llmbs} & \textcolor{black}{Direct optimizer}  & Instructional + contextual prompting
 & Fine-tuned BS position predictor  & DistilBERT  & CNN-only actor, LLM-only actor, combined CNN+LLM actor \\
\cline{2-7}
& \cite{wang2025uavllm} & \textcolor{black}{Direct optimizer}  & Instructional + contextual + iterative prompting        & Prompt-driven UAV planner
 & GPT-4 Turbo, GPT-4o-mini  & Deep Q network (DQN), optimal theoretical upper bound \\
\hline

\multirow{3}{*}{\shortstack[l]{\textbf{Integrated sensing} \\ \textbf{ and communication}}} 
& \cite{li2025isacllm}   & \textcolor{black}{Direct optimizer}   & Instructional + iterative + contextual prompting    & Prompt-guided solver & GPT-3.5 Turbo, GPT-4.0  &  NSGAII, MOEAD, RVEA, AGE-MOEA, MODEA \\
\hline

\multirow{3}{*}{\shortstack[l]{ \textbf{Communication }\\\textbf{standards}\\ \textbf{understanding}}} 
& \cite{alabbasi2025teleoracle}   & \textcolor{black}{Objective/constraint interpretation support}   & Instructional + contextual + RAG prompting & Retrieval-augmented knowledge assistant
 & Phi-2, GPT-4o, GPT-4o-mini  & Base Phi-2, GPT-4o-mini, TeleQnA datasets \\
\cline{2-7}
& \cite{nikbakht2024tspecllm} & \textcolor{black}{Objective/constraint interpretation support}  & Instructional + contextual + RAG prompting  & Retrieval-augmented knowledge assistant & GPT-4, GPT-3.5, Gemini Pro 1.0  & TSpec-LLM dataset \\
\cline{2-7}
& \cite{shao2024wirelessllm} & \textcolor{black}{Objective/constraint interpretation support} & Instructional + Contextual + RAG prompting	  & Retrieval-augmented knowledge assistant	
 & GPT-3.5, GPT-4  & GPT-3.5,GPT-4, wirelessLLM(LLM+RAG) \\
\hline

\multirow{3}{*}{\shortstack[l]{\textbf{Intent processing} \\ \textbf{and network}\\ \textbf{automation}}} 
& \cite{maxenti2025autorAN}    & \textcolor{black}{Automation/control-interface support}    & Instructional + iterative prompting + tool-calling  & Network automation agent & Qwen 2.5 (1.5B-32B)  & Manual configuration \\
\cline{2-7}
& \cite{habib2025intentRAN} & \textcolor{black}{Automation/control-interface support}  & Instructional + contextual + RAG prompting	
 & Fine-tuned user Intent processor	
 & LLaMA (fine-tuned with QLoRA), RAG & Hierarchical Reinforcement learning, Vanilla Decision Transformer \\
\cline{2-7}
& \cite{wu2025oranagent} & \textcolor{black}{Direct optimizer}   & Instructional + contextual + iterative prompting	& Prompt-driven resource allocator	
 & GPT-4  & Random allocation, Equal provisioning, proportional resource allocation \\
 \cline{2-7}
 
 & \cite{liu2024netllm} & \textcolor{black}{Automation/control-interface support}  & Instructional + contextual + RAG prompting & Instruction-following simulation orchestrator	
 & GPT-3.5-Turbo, GPT-4  & Manually scripted simulation in Sionna, Expert-annotated answer \\
\hline

\multirow{4}{*}{\shortstack[l]{\textbf{Low-Altitude} \\ \textbf{Economy and}\\ \textbf{UAV-Assisted}\\ \textbf{Wireless Networks}}} 
& \cite{llm_isac_lae_yang}   & \textcolor{black}{State/context modeling support}     & 	Instructional + contextual + analytical prompting  & 		LLM as system-level optimizer, reasoning engine, and multi-agent coordinator for ISAC
 & GPT-2 (fine-tuned)  & Conventional optimization (e.g., DRL, heuristic methods)    \\
\cline{2-7}
& \cite{uavllmisac_zheng} & \textcolor{black}{State/context modeling support}  & 	Instructional + contextual prompting
 & 	LLM as feature extractor + knowledge distillation teacher
 & Modified GPT-2 (RFF-LLM), Lite-HRNet & ResNet, LmSC, ShuffleNet-v2  \\
\hline
\end{tabular}
}
\end{table}

\begin{table}[t]
\centering
\caption{\textcolor{black}{Optimization-oriented synthesis of LLM-enabled wireless optimization studies.}}
\label{tab:optimization_oriented_summary}
\renewcommand{\arraystretch}{1.02}
\scriptsize
\setlength{\tabcolsep}{2pt}
\begin{tabularx}{\textwidth}{|P{1.75cm}|P{2.0cm}|Y|Y|Y|P{2.0cm}|P{1.35cm}|}
\hline
\textbf{Functional area} 
& \textbf{Representative works} 
& \textbf{Objective / goal} 
& \textbf{Representative constraints} 
& \textbf{Verifier / evaluation} 
& \textbf{Main metrics} 
& \textbf{Timescale} \\
\hline

\textbf{Power control} 
& \cite{zhou2024powercontrol,lee2024llmresource,shao2024wirelessllm,zhou2025promptengineering}
& Maximize rate, EE, SE, or QoS; minimize transmit power
& Power budget, interference, QoS/rate/SINR, RIS/O-RAN limits
& Simulation, external rate/SINR evaluation, DRL/search/baseline comparison
& Sum rate, EE, power, QoS, convergence
& Near-RT / non-RT \\
\hline

\textbf{Bandwidth allocation} 
& \cite{liu2025llmris,noh2025adaptive,wu2025oranagent}
& Maximize rate, fairness, reliability, slice utility, or resource efficiency
& RB/PRB budget, OFDMA assignment, SINR/QoS, queue state, slice limits
& MATLAB/Python feedback, simulation, OAIC/O-RAN testbed
& Rate, fairness, delay, reliability, utility
& Near-RT / non-RT \\
\hline

\textbf{Mobile edge computing} 
& \cite{zhou2025edgecloud,song2024llmmec}
& Minimize service latency; optimize edge--cloud offloading
& Server capacity, wireless/backhaul capacity, RBs, LLM delay, task quality
& Simulation, reward/score feedback, DRL/GA/exhaustive baselines
& Latency, reward, success rate, quality, convergence
& Near-RT / non-RT \\
\hline

\textbf{Traffic prediction and analysis} 
& \cite{zhou2025promptengineering,shokouhi2024multitimespan,wang2024wifi,kan2024mobilellama}
& Predict traffic states and support proactive resource planning
& Historical traffic, spatial--temporal dependence, logs, privacy, data scarcity
& Dataset benchmarks, zero-shot tests, Wi-Fi/network-analysis evaluation
& MAE, accuracy, throughput, delay, code quality
& Offline / non-RT \\
\hline

\textbf{Beam prediction} 
& \cite{sheng2025beam}
& Predict future beam index/direction; maximize beamforming gain
& Mobility, mmWave channel variation, codebook, AoD, prediction horizon
& DeepMIMO benchmark, LSTM baselines, robustness and ablation tests
& Beam gain, accuracy, MSE, robustness
& Real-time / near-RT \\
\hline

\textbf{BS/RIS/UAV deployment} 
& \cite{wang2024bssllm,sevim2024llmbs,wang2025uavllm}
& Improve coverage, received power, user rate, QoS, and placement efficiency
& Site/location limits, coverage radius, traffic demand, blockage, path loss, IAB backhaul
& Real-world data, RAG/agent optimization, Sionna ray tracing, RL baselines
& Coverage, cost, received power, rate, iterations
& Offline / non-RT \\
\hline

\textbf{ISAC} 
& \cite{li2025isacllm}
& Balance communication utility and sensing/localization accuracy
& UAV location, sensing/communication power, CRB, FDMA, area limits
& LEDMA simulation, MOEA baselines, Pareto-front evaluation
& Hypervolume, Pareto quality, CRB, utility
& Non-RT / offline \\
\hline

\textbf{Standards understanding} 
& \cite{alabbasi2025teleoracle,nikbakht2024tspecllm,shao2024wirelessllm}
& Retrieve and interpret telecom standards for constraints and rules
& 3GPP/O-RAN documents, protocol terms, tables, context window, hallucination
& RAG Q\&A, benchmark testing, retrieval/reranking validation
& Accuracy, faithfulness, retrieval quality, hallucination reduction
& Offline / non-RT \\
\hline

\textbf{Intent/network automation} 
& \cite{maxenti2025autorAN,habib2025intentRAN,wu2025oranagent,liu2024netllm}
& Translate intents into policies, xApps/rApps, configurations, or simulation commands
& O-RAN/RIC interfaces, PRBs, QoS/SLA, telemetry, app conflicts, templates
& OAIC/O-RAN testbed, telemetry loop, OpenShift deployment, intent benchmarks
& Throughput, reliability, delay, EE, intent accuracy
& Near-RT / non-RT \\
\hline

\textbf{LAE and UAV-assisted ISAC} 
& \cite{llm_isac_lae_yang,uavllmisac_zheng}
& Support UAV localization, ISAC coordination, and UAV identity/security recognition
& 3-D mobility, A2G channels, NLoS, low SNR, URLLC, RF fingerprints, latency
& Simulation, localization benchmarks, real UAV RFF dataset, KD/PPO compression
& Position error, sensing accuracy, ID accuracy, latency, parameters
& Near-RT / non-RT \\
\hline

\end{tabularx}

\vspace{1mm}
\begin{minipage}{0.98\textwidth}
\scriptsize
\textit{Note:} Objectives, constraints, verifiers, and metrics are summarized at the functional-category level. Detailed paper-specific prompting styles, LLM roles, models, and baselines are reported in Table~\ref{tab:llm_functional_summary}.
\end{minipage}
\end{table}

\section{Prompt Engineering for LLM-enabled Wireless Resource Management}
Prompts serve as the interface between user intent and LLM capabilities. Prompts are natural-language instructions or requests that define tasks or questions and are given as input to the LLMs to generate a specific output. Depending on the quality of prompts, LLMs can produce either highly precise, creative, and informative outputs or vague, general, and incorrect outputs. Prompts can be created and optimized using prompt engineering. This section categorizes the prompt styles utilized in recent LLM-assisted wireless research presented in Section \ref{sec:LLM_taxonomy}, and presents actual prompt examples extracted from the works cited in Table ~\ref{tab:llm_functional_summary}.

\subsection{Instructional Prompts}
Instructional prompting is a technique in which an LLM is guided to perform tasks through carefully crafted natural language instructions that consist of the objective function, problem constraints, input and output formats. This method is both flexible and scalable, enabling  the LLMs to solve unseen problems without requiring any training. This method also allows prompt refinement and iterative adjustments which can mitigate the risk of incorrect outputs. 

\begin{figure}[htbp]
    \centering
    \includegraphics[width=0.65\linewidth]{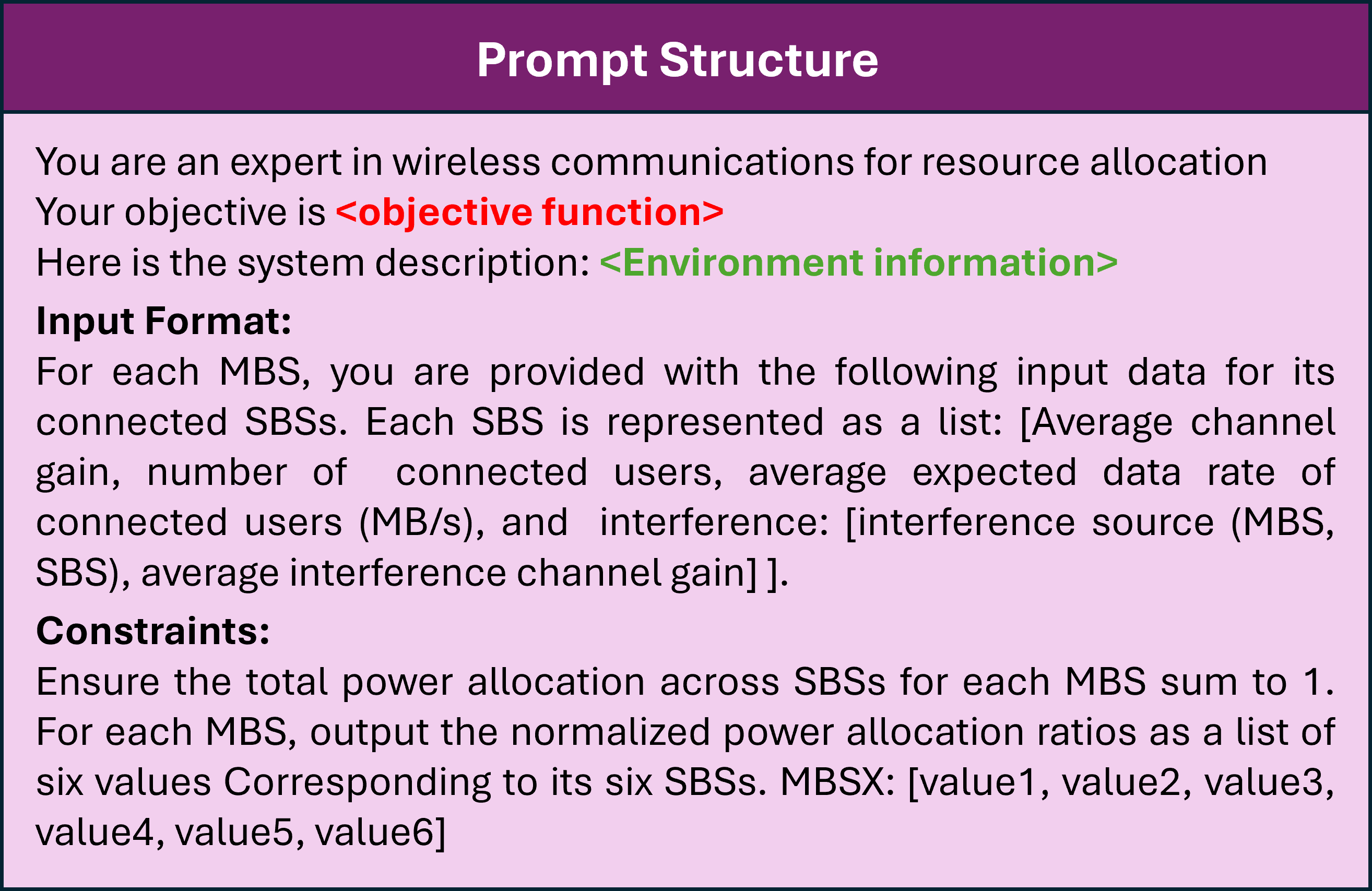}
    \caption{Instructional prompt.}
    \label{fig:ins_llm}
\end{figure}

A representative example is the LLM-hRIC framework proposed in \cite{bao2025llmhric}, where an LLM is deployed within the non-real-time (non-RT) RAN Intelligent Controller to generate strategic resource allocation policies in O-RAN systems. The prompt, as shown in Fig. \ref{fig:ins_llm}, provided to the LLM includes: (i) a clear description of the optimization objective (e.g., maximizing throughput), (ii) environmental information such as the number of macro and small base stations (MBSs/SBSs), channel gains, and interference levels, and (iii) explicit formatting and constraint rules for output power allocations. These structured instructions ensure that the LLM-generated policy complies with system requirements and serves as initial guidance for an RL-based near-RT RIC controller.

\subsection{Few-shot Prompts}
Few-shot prompting involves presenting the LLM with a prompt composed of a few solved examples of the task, along with a new input requiring the same task to be applied. This method enables the model to learn the task pattern or logic without requiring explicit training or parameter updates.

An illustrative example of few-shot prompting was presented in \cite{lee2024llmresource}, where an LLM is used for resource allocation in a wireless system with two interfering transmit–receive pairs. Instead of solving the optimization problem explicitly, the model is prompted with a few example pairs of normalized channel gains and corresponding transmit powers, formatted as simple text (e.g., “If A is ..., then B is ...”). Given a new input, the LLM completes the pattern by generating the power allocation. First, the optimal resource allocation strategy is obtained for a set of example channel conditions. These input-output pairs are then used to construct a prompt for the LLM. Given a new set of channel gains, the LLM generates the corresponding power allocation, which is subsequently mapped to adjust the transmit power for the system.

\subsection{Chain-of-thought Prompts}
In CoT prompting, an LLM is tasked with solving complex problems by encouraging it to break them down into intermediate reasoning subproblems, rather than directly generating the final answers. The prompt includes explicit intermediate steps which need to be performed in order to solve the overall problem, enabling the model to generate more transparent and logically coherent outputs.

An illustrative example of this approach was presented in the WirelessLLM framework \cite{shao2024wirelessllm}, where CoT prompting is used to address a power allocation problem in an OFDM system. Instead of simply asking for an optimal solution, the input prompt directs the LLM to compute intermediate quantities such as channel-to-noise ratios and progressively apply the water-filling algorithm. The model then produces not just the final power allocation, but also executable MATLAB code for each sub-step.
This structured prompting allows the LLM to decompose the problem, integrate analytical logic, and generate implementable solutions. By combining CoT prompting with external tool usage (e.g., MATLAB), WirelessLLM demonstrates reliable and interpretable performance in wireless optimization tasks without requiring traditional mathematical solvers.

\subsection{Role-based Prompts}
Role-based prompting is a technique in which the LLMs are instructed to respond from the perspective of a domain-specific agent or function. This enhances response relevance and consistency by aligning model behavior, i.e., tone and language, with predefined roles or operational contexts. 

A compelling example is Mobile-LLaMA \cite{kan2024mobilellama}, a 5G-network-specialized LLM fine-tuned using role-aligned instruction–response pairs. In this setup, the LLM is trained to act as a network analyst within the Network Data Analytics Function (NWDAF) framework. Each prompt describes a task (e.g., IP routing analysis or packet inspection) as if assigned to a network operations assistant. Through instruction fine-tuning, the model learns to generate Python code using specialized libraries like PyBGPStream, Scapy, and Pandas, enabling it to analyze routing paths, parse packet traces, and assess network performance.

\subsection{Contextual Prompts}
Contextual prompting refers to the technique of embedding environment-specific metadata and task-related information directly into the prompt. This enables LLMs to better understand the problem and generate more contextually accurate information. This approach does not rely on explicit examples, but rather on clearly framed objective functions, constraints, and system configurations to guide the model’s response.

A representative example is LLM-RAO \cite{noh2025adaptive}, which applies contextual prompting to address complex resource allocation in wireless networks. In this setup, the access point (AP) collects real-time communication parameters such as SINR values, queue states, and QoS requirements from multiple users. These parameters are then synthesized into a meta-prompt, which describes the optimization objective (e.g., maximizing data rate or proportional fairness) along with strict system constraints, such as MIMO mode selection and OFDMA restrictions.

\subsection{Retrieval-augmented Prompts}
Retrieval-Augmented Generation (RAG) is a technique, in which the prompts are augmented with relevant documents or facts retrieved from an external knowledge database before generating a response from the LLM. This enhances the model’s performance by grounding its output in up-to-date factual information.

\begin{figure}[htbp]
    \centering
    \includegraphics[width=0.65\linewidth]{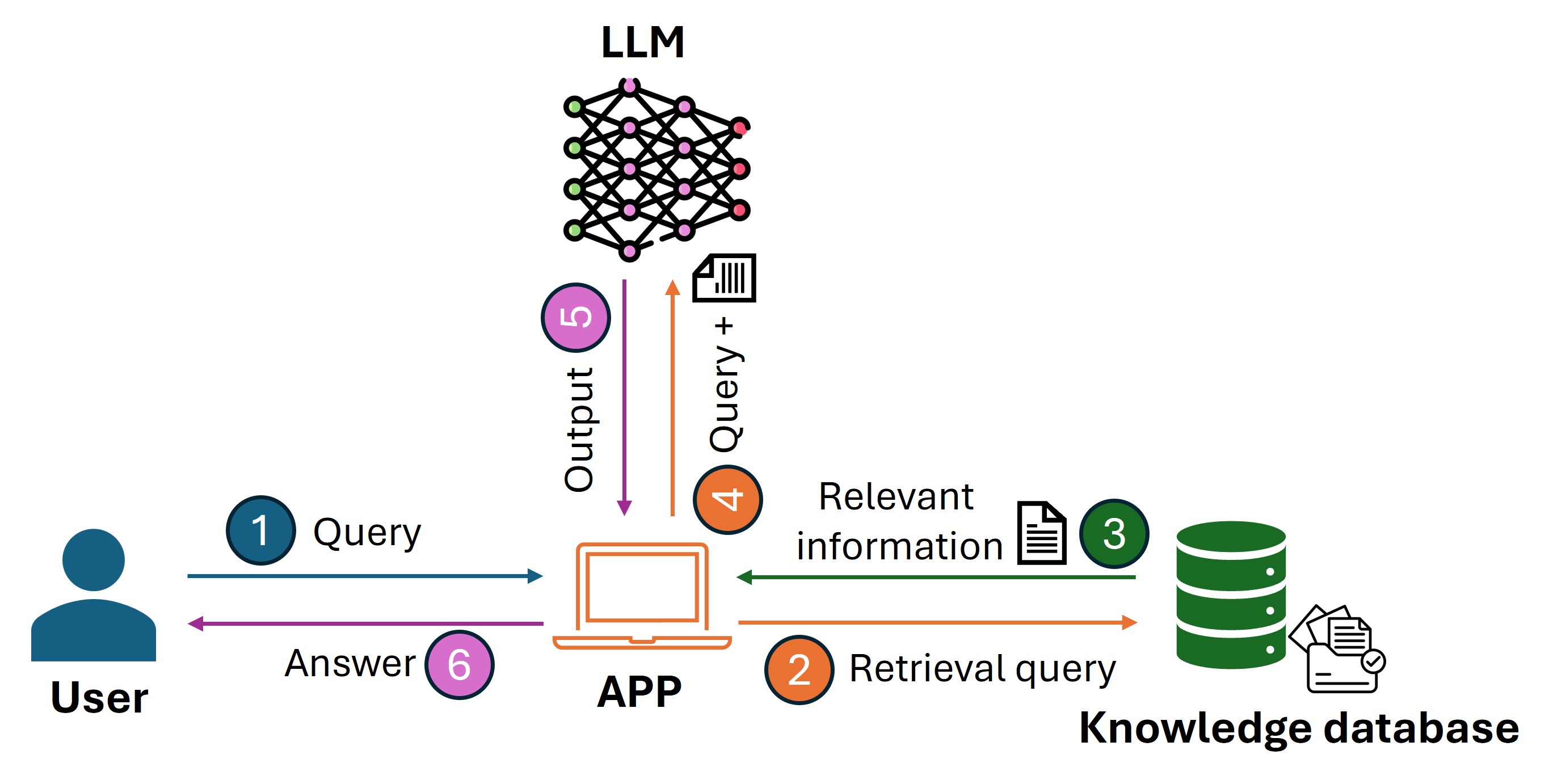}
    \caption{Retrieval-augmented generation.}
    \label{fig:RAG}
\end{figure}

An effective example of this approach was demonstrated in the TSpec-LLM framework \cite{nikbakht2024tspecllm}, which aids LLM understanding of 3GPP telecom specifications. In this setup as shown in Fig. \ref{fig:RAG}, standard documents from 3GPP Releases 8–19 are segmented into smaller chunks, embedded into a vector space, and stored in a vector knowledge database. At inference time, a user query is also embedded and matched to the most relevant document segments. These retrieved text snippets are appended to the query to form an augmented prompt that is then fed into an LLM (e.g., GPT-4 or Gemini) to generate a response.

\subsection{Self-refined Prompts}
In self-refined prompting, the LLM iteratively improves its output by evaluating its performance and incorporating feedback into subsequent prompts. To demonstrate this in the context of wireless networks, a traffic prediction task was considered in \cite{zhou2025promptengineering}, in which an LLM is used to forecast hourly traffic for the next day based on historical data. 

\begin{figure}[htbp]
    \centering
    \includegraphics[width=0.6\linewidth]{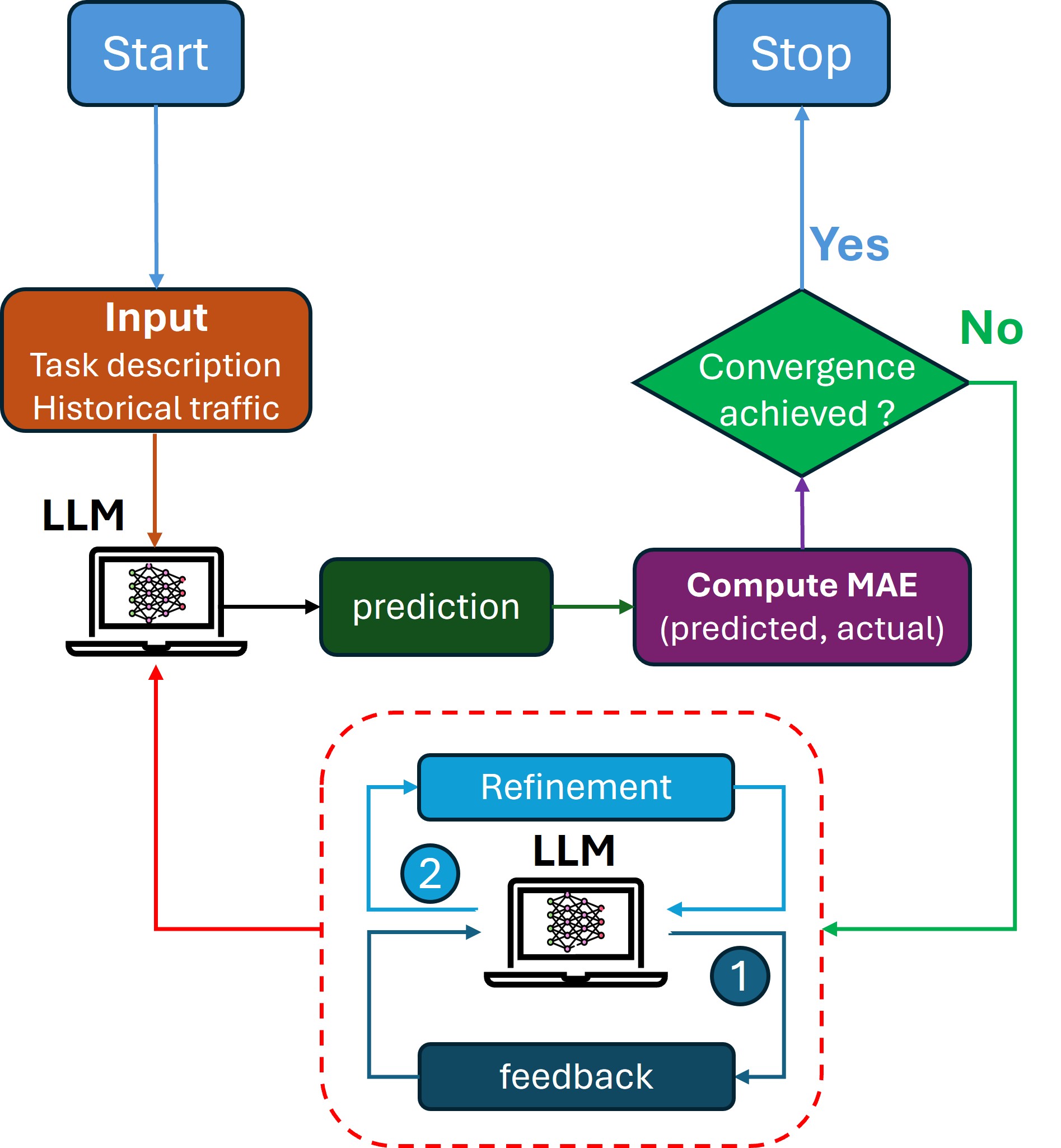}
    \caption{Self-refined prompting.}
    \label{fig:self_refine_flow}
\end{figure}

As illustrated in Fig. \ref{fig:self_refine_flow}, the process starts with an initial prediction generated by the LLM. The output is evaluated using metrics such as mean absolute error, periodicity matching, and formatting completeness. A feedback prompt is then constructed and used to refine the prediction. The loop continues, updating predictions and refining them, until convergence is reached. This self-improving process enables low-complexity, on-device adaptation in dynamic network environments.

\subsection{Prompt-as-prefix}
In this prompting strategy, fixed structured prompts containing domain-specific knowledge, task instructions, or statistical summaries are prepended to the input of a language model to guide its reasoning. Here, contextual understanding is included in the prompt to enhance the model's task-specific understanding and prediction performance.

\begin{figure}[htbp]
    \centering
    \includegraphics[width=0.65\linewidth]{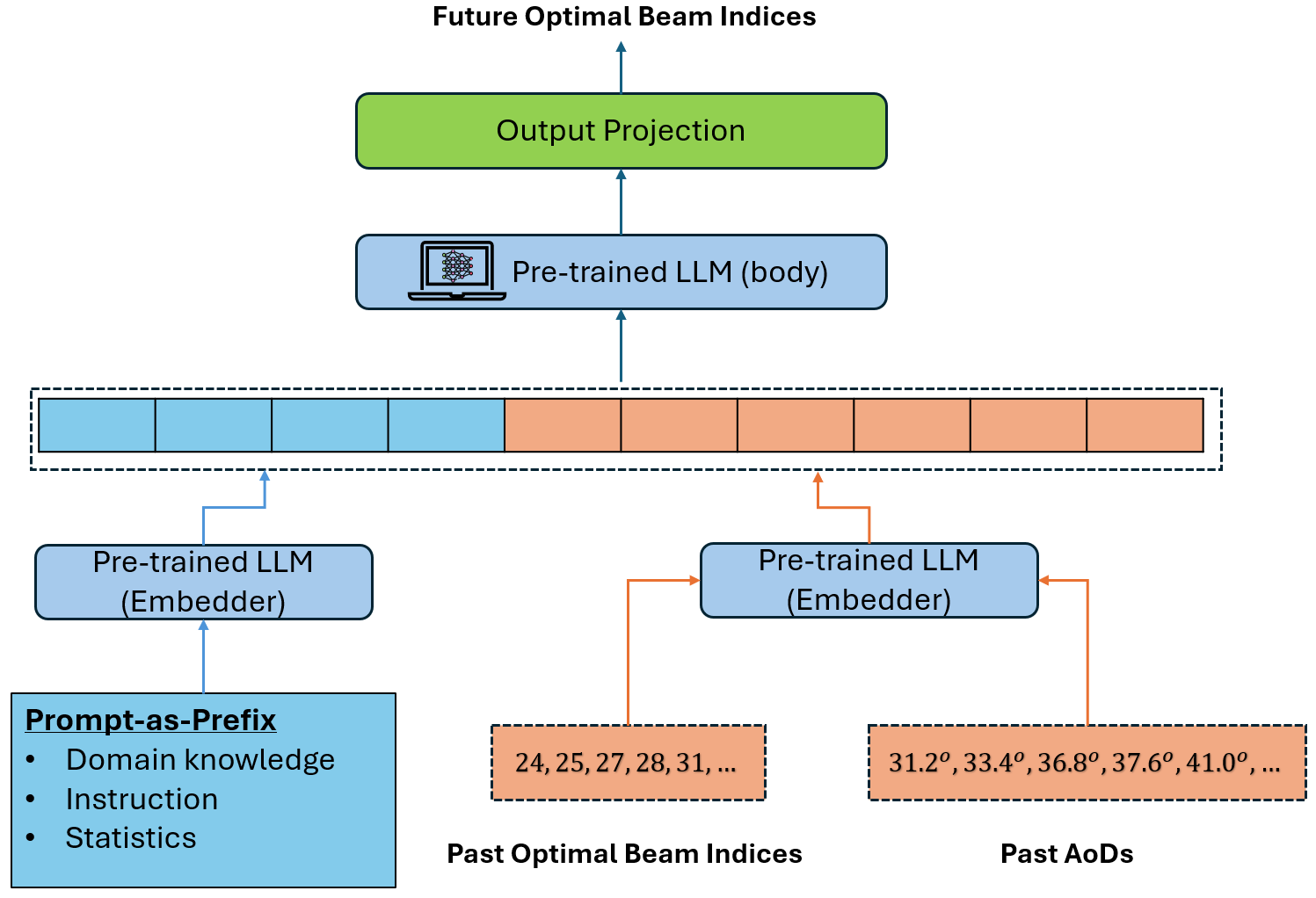}
    \caption{Prompt-as-prefix.}
    \label{fig:Pap_diagram}
\end{figure}

A representative use of PaP in wireless communications was demonstrated in mmWave beam prediction \cite{sheng2025beam}. Historical beam indices and AoDs are converted into patch embeddings and reprogrammed into discrete tokens. To guide the LLM, three types of textual prompts are prepended: (i) Domain Knowledge, describing beam behavior; (ii) Instruction, specifying the prediction task; and (iii) Statistics, summarizing trends and autocorrelations. These prefixes help align the LLM's reasoning with the beam prediction objective.
As shown in Fig.~\ref{fig:Pap_diagram}, this structured prompting improves generalization and robustness across diverse settings, without modifying or fine-tuning the LLM backbone.

{\color{black}
\subsection{Beyond Prompt Engineering: Context and Harness Engineering
}
Beyond individual prompting strategies, recent agentic LLM systems increasingly emphasize context engineering and harness engineering. In LLM-enabled wireless resource management, context engineering refers to the systematic selection, structuring, retrieval, and updating of wireless-domain information provided to the LLM, including CSI, traffic demand, queue states, topology, optimization objectives, constraints, standards documents, historical decisions, and feedback from external evaluators. While prompt engineering focuses on how an instruction is written, context engineering focuses on what task-relevant information is supplied to the model and how it is organized. This is closely related to recent wireless LLM studies that use RAG, prompt-as-prefix conditioning, domain-specific context, or standards-grounded retrieval to improve network reasoning and optimization support~\cite{sheng2025beam,alabbasi2025teleoracle,nikbakht2024tspecllm,shao2024wirelessllm}.

Harness engineering refers to the design of the surrounding execution framework that connects the LLM with external tools, solvers, simulators, verifiers, memory modules, and network-control interfaces. For wireless optimization, this is important because LLM outputs often need to be checked against hard constraints, evaluated by simulators, refined through feedback, or translated into executable network-management actions. Existing LLM-enabled wireless optimization and automation frameworks already follow this direction by integrating LLMs with external evaluators, Python/MATLAB tools, O-RAN controllers, simulators, or closed-loop feedback mechanisms~\cite{noh2025adaptive,wu2025oranagent,liu2024netllm,bao2025llmhric}. Therefore, practical LLM-enabled wireless optimization systems should be viewed not only as prompt-response systems, but also as context-aware and harnessed pipelines that combine LLM reasoning with external validation, simulation, and control mechanisms.
}

\section{Analytical Modeling and Representative Case Studies}
This section moves beyond descriptive review by introducing a generalized analytical model for LLM-driven wireless resource optimization. The proposed model formalizes the role of LLMs as implicit optimizers and provides a common framework for analyzing diverse applications in wireless networks.

\subsection{Generalized Analytical Model for LLM-based Wireless Resource Optimization}
Wireless resource optimization problems studied in the literature, such as scheduling, bandwidth allocation, transmit power control, beam selection, and task offloading in MEC systems, can generally be formulated as constrained optimization problems of the following form,
\begin{align}
\min_{\mathbf{x}} \quad & f(\mathbf{x}; \boldsymbol{\theta}), \\
\text{s.t.} \quad 
& h_i(\mathbf{x}; \boldsymbol{\theta}) \ge 0,\; i = 1,\dots,I, \\
& g_j(\mathbf{x}; \boldsymbol{\theta}) \le 0,\; j = 1,\dots,J.
\end{align}
Here $\mathbf{x}$ denotes the vector of optimization variables, such as transmit power, bandwidth allocation ratios, MEC server indices, and beam indices. Here, $\boldsymbol{\theta}$ represents the system state, including network topology information, user traffic demands, data queue states, wireless channel conditions, and available computational and communication resources. 

The function $f(\mathbf{x};\boldsymbol{\theta})$ represents the objective function, whose specific form depends on the target performance metric considered in the wireless system. Here, the function $f(\cdot)$ is written in minimization form; however, without loss of generality, maximization objectives can be transformed into equivalent minimization problems by minimizing the negative of the objective function. The constraint functions $h_{i}(\cdot)$ and $g_{j}(\cdot)$ capture physical and system limitations commonly captured in wireless networks, such as transmit power budgets, bandwidth budgets, quality-of-service (QoS) requirements, latency constraints, interference limits, and computational capacity constraints at MEC servers. Here $I$ and $J$ are finite non-negative integers representing the numbers of inequality constraints in the optimization problem. 

To solve the above generalized wireless optimization problem using LLMs, we introduce several analytical modeling paradigms that characterize how an LLM can be incorporated into the solution process.

\subsubsection{LLM as a direct solution generator}
In this paradigm, LLMs are utilized as single-step optimizers, generating optimization solutions directly from prompt-encoded problem descriptions. Specifically, the objective function $f(\cdot)$, system state $\boldsymbol{\theta}$, and the constraint functions $\{h_i(\cdot)\}_{i=1}^I$ and $\{g_j(\cdot)\}_{j=1}^J$ are mapped into a structured natural-language representation, denoted by $\textit{Prompt}(\cdot)$, which serves as the input to the LLM. Formally,
\begin{equation}
\begin{aligned}
\textit{Prompt}(\cdot)
&=
\text{Structured natural-language encoding of} \\
&\quad \left(\boldsymbol{\theta}, f(\cdot), \{h_i(\cdot)\}_{i=1}^I,\{g_j(\cdot)\}_{j=1}^J \right).
\end{aligned}
\end{equation}
Let $\mathcal{M}(\cdot)$ denote the inference operation of the LLM which takes the prompt, $\textit{Prompt}(\cdot)$, as input and generates a candidate optimization solution. The resulting solution can be given as 
\begin{equation}
\hat{\mathbf{x}} = \mathcal{M}\!\left(Prompt\!\left(\boldsymbol{\theta}, f(\cdot), \{h_i (\cdot) \}_{i=1}^I, \{g_j(\cdot) \}_{j=1}^J\right)\right).
\end{equation}

This paradigm enables low-latency and flexible solution generation, as the LLM produces a candidate solution in a single inference step based solely on a natural-language description of the optimization problem, without requiring model retraining. As a result, it is well-suited for wireless optimization scenarios with rapidly changing objectives or system conditions. However, since the LLM generates outputs in a probabilistic manner and may exhibit hallucinations, the resulting solution can violate problem constraints, and its optimality is not guaranteed. Consequently, this paradigm is more suitable for optimization problems with soft constraints or scenarios where approximate solutions are acceptable.

\subsubsection{LLM with an external verifier in a closed-loop optimizer}
In this method, LLMs are embedded within a closed-loop optimization framework that incorporates an external verifier to enforce constraint satisfaction and improve solution optimality. The closed-loop optimization process can be executed for a fixed number of iterations or halted early when no additional improvement in solution is observed.

Initially, at iteration $t=0$, let $\boldsymbol{\theta}$ denote the system state and initialize the history set $\mathcal{H}^{(-1)}=\emptyset$. Given prompt encoding of the objective function, system state, constraints, and prior history, the LLM generates  a solution
\begin{equation}
\hat{\mathbf{x}}^{(t)} = \mathcal{M}\!\left(Prompt\!\left(\boldsymbol{\theta}, f(\cdot), \{h_i\}_{i=1}^I, \{g_j\}_{j=1}^{J}, \mathcal{H}^{(t-1)}\right)\right).
\end{equation}
Based on the generated solution $\hat{\mathbf{x}}^{(t)}$, the objective function value $f(\hat{\mathbf{x}}^{(t)};\boldsymbol{\theta})$ and the corresponding constraint violation vector $\mathbf{v}^{(t)}$ are evaluated. The constraint vector $\mathbf{v}^{(t)} \in \mathbb{R}^{I+J}$ quantifies the degree to which the generated solution violates the inequality constraints and is defined as
\begin{equation}
\mathbf{v}^{(t)} =
\begin{bmatrix}
\mathbf{v}_{h}^{(t)} \\
\mathbf{v}_{g}^{(t)}
\end{bmatrix},
\end{equation}
where the individual components are given by
\begin{align}
v_{h_i}^{(t)} &= \max\left(0,-h_i(\hat{\mathbf{x}}^{(t)};\boldsymbol{\theta})\right),
\quad i = 1,\ldots,I, \\
v_{g_j}^{(t)} &= \max \left(0, g_j(\hat{\mathbf{x}}^{(t)};\boldsymbol{\theta})\right),
\quad j = 1,\ldots,J.
\end{align}
The tuple $\left( \hat{\mathbf{x}}^{(t)}, f(\hat{\mathbf{x}}^{(t)};\boldsymbol{\theta}), \mathbf{v}^{(t)} \right)$ is appended to the history set and the updated history is given as 
\begin{equation}
    \mathcal{H}^{(t)}=\mathcal{H}^{(t-1)} \cup \Big\{ \left( \hat{\mathbf{x}}^{(t)}, f(\hat{\mathbf{x}}^{(t)};\boldsymbol{\theta}), \mathbf{v}^{(t)} \right) \Big\}.
\end{equation}
The updated history $\mathcal{H}^{(t)}$ is embedded into the subsequent prompt, enabling the LLM to generate an improved solution at iteration $t+1$ as
\begin{equation}\label{eq:nprompt}
\hat{\mathbf{x}}^{(t+1)} = \mathcal{M}\!\left(Prompt\!\left(\boldsymbol{\theta}, f(\cdot), \{h_i\}_{i=1}^I, \{g_j\}_{j=1}^{J}, \mathcal{H}^{(t)}\right)\right).
\end{equation}
The process is repeated until a stopping criterion, such as convergence or a maximum number of iterations, is satisfied.
This closed-loop paradigm is particularly effective for complex wireless optimization problems involving mixed-integer variables, tightly coupled constraints, and dynamically changing objectives.

\subsubsection{LLM as a representation encoder in a trainable neural pipeline}
In this paradigm, LLMs are not directly employed to generate optimization solutions. Instead, a pretrained and typically frozen LLM is embedded within a trainable neural architecture to extract high-level features from structured natural-language representations of the objective function, system state, constraint functions, and prior solutions.

Let $\boldsymbol{\phi}_{\text{pre}}(\cdot;\boldsymbol{\alpha})$ denote the trainable encoder network placed before the LLM. It takes as input the system state $\boldsymbol{\theta}$ together with structured natural-language instructions $\mathcal{I}$, and produces an intermediate representation suitable for the LLM input space,
\begin{equation}
    \mathbf{Z}=\boldsymbol{\phi}_{\text{pre}}(\boldsymbol{\theta},\mathcal{I};\boldsymbol{\alpha}).
\end{equation}
The pretrained LLM $\mathcal{M}(\cdot)$, with frozen parameters, then transforms this representation into a latent embedding
\begin{equation}
    \mathbf{E}=\mathcal{M}(\mathbf{Z}).
\end{equation}
Finally, a trainable decoder network $\boldsymbol{\phi}_{\text{post}}(\cdot;\boldsymbol{\beta})$ maps the LLM-generated embedding to the optimization variables
\begin{equation}
    \hat{\mathbf{x}}=\boldsymbol{\phi}_{\text{post}}(\mathbf{E};\boldsymbol{\beta}).
\end{equation}
Here, $\hat{\mathbf{x}}$ denotes the predicted resource allocation or decision vector.
Let $\boldsymbol{\phi} = (\boldsymbol{\phi}_{\text{pre}}, \boldsymbol{\phi}_{\text{post}})$ denote the collection of trainable neural network components surrounding the LLM. Given a supervised learning training dataset $\{\boldsymbol{\theta}^{(n)}, \bar{\mathbf{x}}^{(n)}\}_{n=1}^{N}$, where $\boldsymbol{\theta}^{(n)}$ denotes the system state and $\bar{\mathbf{x}}^{(n)}$ represents the corresponding target solution (e.g., optimal or near-optimal), the model can be trained using supervised learning.
Specifically, the parameters $(\boldsymbol{\alpha}, \boldsymbol{\beta})$ can be learned by minimizing the mean squared error (MSE) loss
\begin{align}
&\mathcal{L}(\boldsymbol{\alpha}, \boldsymbol{\beta})
\nonumber\\ 
&=\frac{1}{N}
\sum_{n=1}^{N}
\left\|
\boldsymbol{\phi}_{\text{post}}\!\left(
\mathcal{M}\!\left(
\boldsymbol{\phi}_{\text{pre}}(\boldsymbol{\theta}^{(n)}, \mathcal{I}^{(n)}; \boldsymbol{\alpha})
\right);
\boldsymbol{\beta}
\right)
-
\bar{\mathbf{x}}^{(n)}
\right\|_2^2 .
\end{align}
During training, gradients are propagated only through the trainable components $\boldsymbol{\phi}$, while the parameters of the pretrained LLM $\mathcal{M}(\cdot)$ remain fixed.

\subsection{Case Study I: LLM-based Beam Prediction for Millimeter Wave Systems}
In this study, we analyze a representative LLM-based beam prediction framework proposed in \cite{sheng2025beam}, which illustrates how LLMs can be utilized to learn high-level representations for  complex wireless optimization problems.
Specifically, a downlink millimeter-wave (mmWave) beam prediction scenario is considered, where a single-antenna user communicates with a BS equipped with a uniform linear array (ULA) comprising $M$ antennas. The user is mobile, which necessitates accurate beam prediction under time-varying channel conditions. The channel between the user and the BS is characterized using the Saleh-Valenzuela (SV) model.
The BS consists of a single radio-frequency chain and a phase-shifter-based analog beamforming architecture. The analog beamformer employs a Discrete Fourier Transform (DFT) codebook $\mathcal{F}$ consisting of $Q$ candidate beams. 

The objective is to predict the optimal beam directions for the upcoming $H$ time steps based on historical observations from the preceding $U$ time steps, including the optimal beam and corresponding AoDs. Mathematically, the problem is expressed as follows
\begin{equation}
    \max_{\{ \hat{q}^{*}_{1},...,\hat{q}^{*}_{k},...  \}} \sum_{k=1}^{H}\frac{|\mathbf{h}_{k}^{\mathrm{T}}\mathbf{f}^{(\hat{q}^{*}_{k})}|^{2}}{|\mathbf{h}_{k}^{\mathrm{T}}\mathbf{f}^{(q_{k}^{*})}|^{2}},
\end{equation}
where $\mathbf{h}_{k}$ denotes the channel vector at the $k$-th time slot, and $\hat{q}_{k}^{*} \in \{0,1,2,...,Q-1 \}$ represents the predicted beam index. The corresponding candidate transmit beam is given by $\mathbf{f}^{(\hat{q}_{k}^{*})}$. The optimal beam index at time slot $k$, denoted by $q_{k}^{*}$, is given by
\begin{equation}
    q_{k}^{*}=\arg \max_{q_{k} \in \{0,1,2,...,Q-1\}} \left| \mathbf{h}_{k}^{T} \mathbf{f}^{(q_{k})} \right|.
\end{equation}

The solution framework for this problem falls under the representation-based learning paradigm, in which a pretrained and frozen LLM is embedded within a trainable neural pipeline. The surrounding neural components are trained using supervised learning. The DeepMIMO dataset \cite{ahmedDeepMIMO}, generated using precise ray-tracing techniques, provides the beam indices and AoDs information for training.
Each sample $n$ in the dataset consists of an input sequence, denoted as $\mathbf{X}^{(n)}=\{ \mathbf{q}_{t-U+1:t}, \boldsymbol{\varphi}_{t-U+1:t} \}$ which includes the optimal beam indices and corresponding AoDs over the previous $U$ time steps, and an output label $\mathbf{Y}^{(n)}=\{ \mathbf{q}_{t+1:t+H}\}$ representing the optimal beam indices over the subsequent $H$ time steps.

The pre-LLM encoder $\boldsymbol{\phi}_{\mathrm{pre}}(\cdot;\boldsymbol{\alpha})$ processes the historical beam index and AoD sequences and fuses them into a common latent representation using  attention-based modules. The resulting output is a sequence of latent tokens, i.e., fixed-dimensional embedding vectors representing aggregated temporal patches of the input observations, denoted by $\mathbf{Z}_{\mathrm{data}}\in \mathbb{R}^{T_{d}\times D}$, where $T_{d}$ denotes the number of tokens and $D$ corresponds to the LLM hidden dimension.
In addition, a structured natural-language prompt encoding task instructions and system context is embedded as a prefix to the data token sequence, denoted by $\mathbf{Z}_{\mathrm{prompt}}\in \mathbb{R}^{T_{p} \times D}$, where $T_p$ represents the number of prompt tokens.
The final LLM input sequence is constructed by concatenating the prompt tokens and data tokens as
\begin{equation}
\mathbf{Z} = \big[ \mathbf{Z}_{\mathrm{prompt}}; \mathbf{Z}_{\mathrm{data}} \big]
\in \mathbb{R}^{(T_p + T_d)\times D}.
\end{equation}
Accordingly, the pre-LLM encoder can be abstracted as
\begin{equation}
\mathbf{Z} = \boldsymbol{\phi}_{\mathrm{pre}}(\boldsymbol{\theta}, \mathcal{I}; \boldsymbol{\alpha}),
\end{equation}
where $\boldsymbol{\theta}$ denotes the system state and $\mathcal{I}$ represents the structured prompt. The pretrained LLM then transforms the input token sequence into a sequence of hidden representations
\begin{equation}
\mathbf{E} = \mathcal{M}\big(\mathbf{Z}\big) \in \mathbb{R}^{(T_p + T_d)\times D}.
\end{equation}
The post-LLM decoder 
$\boldsymbol{\phi}_{\mathrm{post}}(\cdot;\boldsymbol{\beta})$ is implemented as a lightweight linear projection that maps the LLM output embeddings to the predicted beam indices. Specifically, given the LLM hidden representations 
$\mathbf{E}$, the decoder produces the prediction
\begin{equation}
\hat{\mathbf{Y}} = \boldsymbol{\phi}_{\text{post}}(\mathbf{E}; \mathbf{W},\mathbf{b}) = \mathbf{E}\mathbf{W} + \mathbf{b},
\end{equation}
where $\mathbf{W}$ and $\mathbf{b}$ denote trainable parameters, and $\hat{\mathbf{Y}}$ represents the predicted beam index sequence over the future horizon.

The model is trained using supervised learning by minimizing the mean squared error (MSE) between the predicted and ground-truth beam indices over the H-step prediction horizon
\begin{equation}
\mathcal{L}=\frac{1}{N}\sum_{n=1}^{N}\frac{1}{H}
\sum_{k=1}^{H}
\left\|
\hat{\mathbf{Y}}_{k}^{(n)} - \mathbf{Y}_{k}^{(n)}
\right\|_2^2.
\end{equation}
Here, $\hat{\mathbf{Y}}$ denotes the predicted beam index vector and $\mathbf{Y}$ denotes the corresponding ground-truth optimal beam index vector.

\begin{figure}[t]
    \centering
    \includegraphics[width=0.75\linewidth]{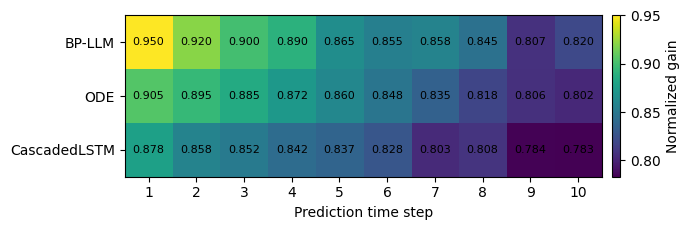}
    \caption{\textcolor{black}{Normalized beamforming gain versus prediction horizon for LLM-based and LSTM-based beam prediction methods. The results are author-generated simulation results obtained by implementing the beam-prediction setting and baseline methods following \cite{sheng2025beam}.}}
    \label{fig:bp_case}
\end{figure}

In order to evaluate the performance of the LLM-based beam prediction scheme, the original study adopts the normalized beamforming gain as the primary performance metric, which reflects the accuracy of predicted beams relative to the optimal beam. Fig. \ref{fig:bp_case} shows the normalized beamforming gain achieved by the LLM-based predictor and representative LSTM-based baselines as a function of the prediction horizon, based on simulations using the DeepMIMO dataset.
As illustrated in Fig. \ref{fig:bp_case}, the LLM-based predictor consistently outperforms LSTM-based models, with the performance gap widening for longer prediction horizons. This trend indicates a stronger ability to capture long-term temporal dependencies in beam dynamics, highlighting the robustness of LLMs as high-capacity representation learners for beam prediction in dynamic mmWave environments.

In addition to prediction accuracy, we evaluate the computational overhead of the considered models during online inference. {\color{black}Table~\ref{tab:beam_inference_overhead} reports author-generated inference measurements for the BP-LLM and Cascaded LSTM models, including inference latency, peak GPU memory consumption, average power usage, and energy cost per inference.} The results show that the Cascaded LSTM model achieves lower latency and a much smaller memory footprint, while the BP-LLM model still maintains acceptable inference latency relative to the 16 ms prediction interval. These results indicate that the LLM-based predictor, despite its larger memory requirement, can be practical for real-time beam prediction when deployed with suitable hardware support.

\begin{table}[t]
\centering
\caption{Inference-overhead comparison of BP-LLM and Cascaded LSTM for the beam prediction task.}
\label{tab:beam_inference_overhead}
\setlength{\tabcolsep}{6pt}
\renewcommand{\arraystretch}{1.1}
\begin{tabular}{|c|c|c|}
\hline
\textbf{Measurement} & \textbf{BP-LLM} & \textbf{Cascaded LSTM} \\
\hline
Avg. latency (ms) & 4.73 & 1.74 \\
\hline
Peak memory (MB) & 949.67 & 24.73 \\
\hline
Avg. power (W) & 22.55 & 37.41 \\
\hline
Energy / inference (J) & 0.137 & 0.054 \\
\hline
\end{tabular}
\end{table}

\subsection{Case Study II: LLM-based Adaptive Resource Allocation}
In this study, we analyze an LLM-based resource allocation optimizer (LLM-RAO) proposed in \cite{noh2025adaptive}, which leverages an LLM to jointly optimize OFDMA resource allocation, single-user and multi-user MIMO mode selection, and MU-MIMO user scheduling, subject to QoS constraints and queuing models in dynamic wireless environments. 
An IEEE 802.11ax uplink scenario is considered, in which a single access point (AP), equipped with $M$ antennas, serves $K$ single-antenna users using an OFDMA-based multiple access scheme. The AP is responsible for estimating uplink channels and allocating resource blocks (RBs) to one or more users, and initiating uplink transmissions. 
Each RB is indexed by a tuple $(l,i)$, where $l$ denotes the RB level and $i$ represents the corresponding index.
The data rate of user $k$ on $\mathrm{RB}_{l,i}$ is given by $r_{l,i}^{(k)}=B_{l,i}\log(1+\Gamma_{l,i}^{(k)})$ (bits/s), where $B_{l,i}$ denotes the bandwidth of $\mathrm{RB}_{l,i}$. Each user maintains a backlog of packets, while new packet arrivals are modeled as a Poisson process.

The goal of the resource optimization is to maximize the network's data rate while satisfying RB allocation constraints, queue dynamics, and QoS guarantees. Accordingly, the optimization problem can be formulated as follows
\begin{align}
\max_{\mathbf{X}} \quad 
& \sum_{k \in \mathcal{K}} r^{(k)}, \label{eq:rao_obj} \\
\text{s.t.} \quad
& \sum_{\forall (l,i)}x_{l,i}^{(k)}\leq 1, \quad \forall k \in \mathcal{K},\\
& \sum_{\forall (l,i)} \delta \left( \sum_{k \in \mathcal{K}}x_{l,i}^{(k)}\mathbf{e}_{l,i} \preceq \mathbf{1}_{q} \right ),\\
& \sum_{k \in \mathcal{K}}x_{l,i}^{(k)}\leq G(l), \quad \forall (l,i),\\
& \sum_{\forall (l,i)} r_{l,i}^{(k)} \geq x_{l,i}^{(k)}R^{(k)}, \label{eq:rao_c2}
\end{align}
where $x_{l,i}^{(k)}=1$, if the $k$-th user is allocated to $\mathrm{RB}_{l,i}$ and $x_{l,i}^{(k)}=0$ otherwise. $\mathbf{e}_{l,i} \in \mathbb{Z}^{q\times 1}$ is an indicator vector and $q$ denotes the number of $26$-tone RBs. $R^{(k)}$ denotes the minimum rate requirement of user $k$.

The proposed solution method for the above problem employs an LLM as a direct optimizer embedded in a closed-loop inference framework. At each scheduling interval, the AP first generates a structured natural-language prompt that encodes the current system state, including the  objective function $\sum_{k\in \mathcal{K}} r^{(k)}$, users' SINR values $\boldsymbol{\Gamma}$, queue backlogs $\mathbf{Q}$, QoS requirements $\mathbf{R}$, and resource allocation constraints $\mathcal{C}$. The prompt is given as input to the LLM, which generates a solution as
\begin{equation}
\hat{\mathbf{x}}^{(0)} =
\mathcal{M}\!\left(
Prompt\!\left(
\boldsymbol{\Gamma},
u(\cdot), \mathbf{Q}, \mathbf{R}, \mathbf{\mathcal{C}},
\mathcal{H}^{(-1)}
\right)
\right),
\end{equation}
where $\mathcal{H}^{(-1)}=\emptyset$. The generated solution $\hat{\mathbf{x}}^{(0)}$ is then evaluated using an 
external toolkit (MATLAB) according to the user-wise
data-rate $\{r^{(k)}\}_{k \in \mathcal{K}}$. Based on this
evaluation, the aggregate optimization score
$r(\hat{\mathbf{x}}^{(0)}) = \sum_{k \in \mathcal{K}} r^{(k)}$ is computed.
In addition, constraint satisfaction is explicitly verified, and any
violations are recorded as negative feedback.
The evaluation results are incorporated into the solution history, i.e., $\mathcal{H}^{(0)}$, by
appending the tuple
$\left(\hat{\mathbf{x}}^{(0)}, r(\hat{\mathbf{x}}^{(0)}), \mathbf{v}^{(0)}\right)$,
where $\mathbf{v}^{(t)}$ denotes the constraint violation vector. This
updated history is embedded into the subsequent prompt as in (\ref{eq:nprompt}) and provided to
the LLM, enabling it to generate an improved solution through
in-context learning.
The above closed-loop inference process is repeated iteratively until
the LLM can no longer propose solutions with improved optimization
scores or until a predefined maximum number of optimization steps is
reached.

\begin{figure}[htbp]
    \centering
    \includegraphics[width=0.75\linewidth]{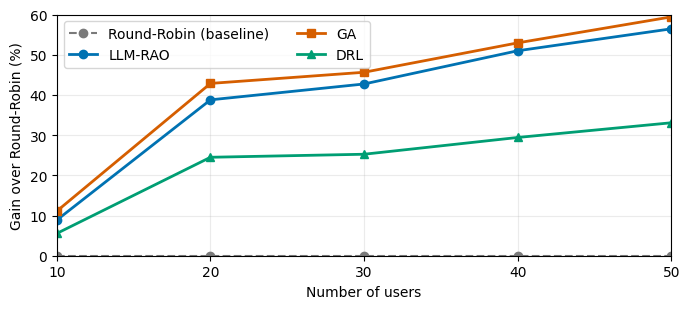}
    \caption{\textcolor{black}{Aggregate data rate versus number of users for different resource-allocation strategies. The results are author-generated simulation results obtained using the simulation setting and benchmark methods following \cite{noh2025adaptive}.}}
    \label{fig:ra_case}
\end{figure}

The original study evaluates the LLM-based resource-allocation scheme using aggregate system data rate as the main metric. Fig.~\ref{fig:ra_case} compares the LLM-based optimizer with Round-Robin scheduling, DRL, and a genetic algorithm-based near-optimal benchmark as the number of users increases. The learning- and optimization-based schemes substantially outperform Round-Robin, with larger gains at higher user densities. The LLM-based optimizer achieves about $10$--$55\%$ improvement and closely follows the genetic algorithm, which reaches about $12$--$60\%$, while DRL provides smaller gains of roughly $5$--$32\%$. This trend suggests that prompt-driven, closed-loop LLM optimization can better adapt to combinatorial and constrained resource-allocation settings than fixed-policy learning approaches.

\begin{figure}[htbp]
    \centering
    \includegraphics[width=0.75\linewidth]{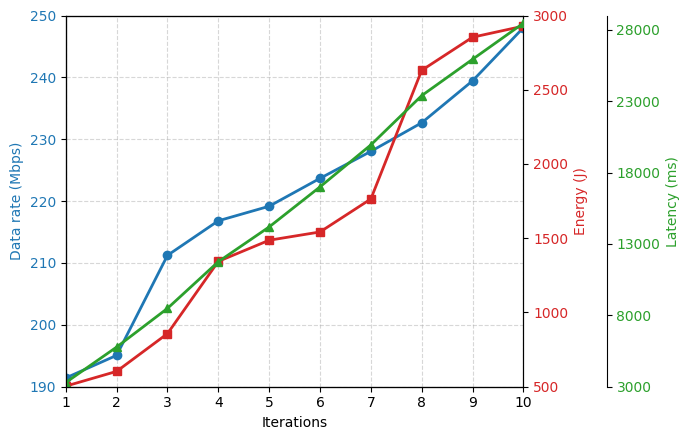}
    \caption{Performance--energy--latency trade-off vs iterations.}
    \label{fig:ra_energy_rate_latency}
\end{figure}

To estimate the computational overhead of LLM-based resource allocation, we implemented a local prototype using Qwen2.5-Coder-7B on an NVIDIA RTX 4090 GPU with 24 GB memory. The LLM generates candidate resource-allocation solutions, which are evaluated by an external Python module that measures optimization latency, GPU memory usage, power consumption, and energy consumption using NVML. The peak GPU memory usage reached 7786 MB, and the average GPU power consumption was 142.89 W. {\color{black}Fig.~\ref{fig:ra_energy_rate_latency} shows the author-generated performance--energy--latency trade-off for $K=20$ under different iteration budgets.} Additional iterations improve the achieved rate but significantly increase latency and energy consumption. Thus, iterative LLM-based resource allocation is currently unsuitable for fine-grained real-time wireless control, but remains promising for slower-timescale tasks such as supervisory optimization, policy adaptation, and offline resource management.

\begin{table}[htbp]
\centering
\caption{\textcolor{black}{Author-generated BSS optimization inference results for Gemma2:9B variants under different quantization levels.}}
\label{tab:gemma2_quantization}
\begin{tabular}{|c|c|c|c|}
\hline
\textbf{Metric} & \textbf{FP16} & \textbf{Q8\_0} & \textbf{Q4\_0} \\
\hline
Valid output (\%) & 100.0 & 94 & 89 \\
\hline
Normalized cost & 12.24 & 11.98 & 11.85 \\
\hline 
Coverage (\%) & 96.68 & 94.31 & 93.50 \\
\hline
Latency (ms) & 1685.17 & 1188.02 & 936.68 \\
\hline
Energy (J) & 416.75 & 283.32 & 211.14 \\
\hline
Peak GPU memory (MB) & 23313.18 & 14929.18 & 10747.18 \\
\hline
\end{tabular}
\end{table}

{\color{black}
\subsection{Deployment Feasibility of Efficient LLM Inference for Wireless Optimization}
A key challenge in LLM-enabled wireless optimization is that LLM inference must be aligned with the timescale of the target optimization loop. Real-time PHY-layer functions, such as scheduling, link adaptation, beam tracking, and fast power control, may require sub-millisecond to millisecond responses, whereas near-RT RIC control, slicing, traffic steering, and edge offloading can tolerate tens of milliseconds to one second. Non-RT policy generation, intent interpretation, standards reasoning, and offline deployment planning can operate on longer timescales. Therefore, the relevance of efficient LLM deployment should be evaluated not as a general LLM-system benchmark, but in terms of whether latency, memory, and energy consumption allow the LLM to be placed in the appropriate wireless optimization layer. This subsection discusses model compression, hierarchical control, hardware acceleration, and edge/distributed inference as practical techniques \cite{Eff_llm_lirui}, for matching LLM capabilities with wireless optimization timescales.
}

\subsubsection{Model compression}
Model compression reduces the computational and memory overhead of LLM inference, making LLMs more suitable for latency-sensitive and resource-constrained wireless environments. Representative techniques include quantization, pruning, and knowledge distillation.

Quantization lowers the numerical precision of model parameters and activations, e.g., from FP32 to FP16, INT8, or INT4, thereby reducing memory usage, memory bandwidth, and inference latency. Representative methods include LLM.int8~\cite{NEURIPS2022_llm8}, SmoothQuant~\cite{pmlr_smoothquant}, and GPTQ~\cite{frantar_gptq}. 
{\color{black}To illustrate wireless-optimization-specific deployment feasibility, we evaluate Gemma2:9B variants under FP16, Q8\_0, and Q4\_0 quantization on an author-generated base-station sitting (BSS) prompt benchmark inspired by~\cite{wang2024bssllm}. The LLM generates structured deployment decisions from candidate sites, weak-coverage demand points, deployment costs, coverage radii, distance constraints, and a traffic-coverage requirement; these decisions are then checked and repaired by an external verifier. As shown in Table~\ref{tab:gemma2_quantization}, quantization reduces latency, energy consumption, and peak GPU memory while maintaining valid outputs and traffic coverage above the required threshold, supporting verifier-assisted near-RT or non-RT wireless optimization workflows.}

Pruning removes less important weights, neurons, channels, or attention heads to reduce model size and inference cost. Structured pruning is particularly attractive for hardware deployment, while recent methods such as SparseGPT~\cite{frantar_sparsegpt}, LLM-Pruner~\cite{NEURIPS2023_llm_pruner}, and Wanda~\cite{sun2024s_wanda} can achieve high sparsity with limited retraining.

Knowledge distillation transfers knowledge from a large teacher model to a smaller student model, enabling lower-cost inference with reduced memory and latency. Black-box, chain-of-thought~\cite{cheng_y_cot_distill}, and instruction distillation~\cite{jiang_instditill} are commonly used to improve compact models, although they may require additional data generation or training overhead.

\subsubsection{Hierarchical control architectures}
A promising direction is to integrate LLMs into hierarchical control frameworks that partition control across timescales. LLMs operate at higher layers, typically in the non-RT RIC ($>1$s), performing global analysis, intent interpretation, and long-term policy optimization. Latency-critical functions in the near-RT RIC (10 ms-1s) and lower layers are handled by lightweight controllers and specialized learning algorithms.

Recent works demonstrate this mapping. LLM-based RAN frameworks place LLMs in the non-RT RIC to process network-wide telemetry and generate structured policies delivered via the A1 interface \cite{oluwaseyi2025the}. Similarly, LLM-hRIC treats the non-RT RIC as a strategic layer that produces high-level policies (e.g., power allocation, slicing), which are refined by RL-based xApps using real-time E2 measurements \cite{bao2025llmhric}. Agent-based architectures further extend this hierarchy across rApps, xApps, and lower layers for coordinated multi-timescale control \cite{elkael2026agentranagenticaiarchitecture}.

In this paradigm, the LLM acts as a slow-timescale policy generator, producing configurations such as scheduling priorities or slicing objectives. These are translated into constraints, reward signals, or parameters and passed to near-RT controllers via standardized interfaces (e.g., A1). Fast control loops, such as scheduling, link adaptation, and power control, are then executed locally using lightweight methods. This separation removes LLM inference from latency-critical paths while enabling global, context-aware optimization.

\begin{table}[htbp]
\centering
\caption{\textcolor{black}{Author-generated CPU/GPU inference-efficiency comparison for Gemma2:9B on the BSS optimization benchmark.}}
\label{tab:cpu_gpu_gemma2_9b}
\begin{tabular}{|c|c|c|c|}
\hline
\textbf{Mode} & \textbf{Coverage}  & \textbf{Latency} & \textbf{Speedup} \\
 & \textbf{(\%)} &   \textbf{(ms)} &  \\
\hline
GPU & 96.13  & 1814.99 & $21.0\times$ \\
\hline
CPU & 96.13  & 38203.16 & $1.0\times$ \\
\hline
\end{tabular}
\end{table}

\subsubsection{Hardware acceleration}

Hardware acceleration is essential for deploying LLMs under latency and energy constraints. LLM inference is dominated by large matrix operations, memory-bandwidth bottlenecks, and sequential autoregressive decoding, which motivates specialized hardware and system-level optimization~\cite{li2025llm_hardwareaccelratsurvey}. GPUs, FPGAs, ASICs, and in-memory computing have been explored to improve throughput and energy efficiency, while hardware-aware techniques such as algorithm--hardware co-design and speculative decoding further reduce inference latency~\cite{koilia2024hardwareaccelerationllmscomprehensive,yang_hades}.

To illustrate the impact of hardware acceleration, we compare CPU and GPU execution of ``gemma2:9b-instruct-fp16'' on the BSS prompt benchmark using an Intel(R) Core(TM) Ultra 9 285K CPU and an NVIDIA RTX 4090 GPU. {\color{black}Building on the BSS quantization results in Table~\ref{tab:gemma2_quantization}, we further evaluate hardware acceleration on the same verifier-assisted BSS optimization benchmark. As shown in Table~\ref{tab:cpu_gpu_gemma2_9b}, GPU inference reduces the average latency from 38203.16 ms to 1814.99 ms, corresponding to an approximately $21.0\times$ speedup, while maintaining the same verifier-repaired traffic coverage and normalized deployment cost. Together, these results show that quantization and hardware acceleration can reduce local LLM inference overhead for near-RT policy generation, verifier-assisted optimization, and edge-based network automation.}

\subsubsection{Edge and distributed inference}
Advances in edge computing and distributed inference provide an effective way to reduce LLM inference latency in wireless networks. By deploying LLM components across edge servers and devices, computation can be performed closer to users, reducing communication delay and alleviating bandwidth constraints.

A key approach is distributed LLM inference, where model layers or tensors are partitioned across multiple nodes. For example, layer-wise deployment distributes transformer layers across heterogeneous edge nodes to balance computation and communication \cite{picano_matchllm}, while inference allocation frameworks optimize task placement across edge servers under resource and latency constraints \cite{zhu_bird}. Tensor parallelism further enables collaborative inference by splitting model parameters across devices \cite{zhang_com_distllminf}. In addition, edge-device collaborative architectures combine lightweight on-device models with edge-hosted LLMs to improve latency, for example through distributed speculative decoding \cite{ning2025dssdefficientedgedevicellm}. Federated-style approaches further enable collaborative inference while reducing communication and preserving data privacy \cite{deng2025federatedattentiondistributedparadigm}.

Overall, edge acceleration and distributed inference enable scalable and low-latency LLM deployment by jointly optimizing computation and communication across edge-cloud infrastructure.

\section{Hybrid LLM-based Approaches in Artificial Intelligence}
\subsection{Hybrid LLM--Machine Learning Approaches} 
Integrating the generative and inferential capabilities of LLMs with the data analysis strengths of ML models establishes a multi-faceted framework that surpasses the performance achievable by either model alone. Within this context, LLMs play several critical roles: (i) translating high-level, context-rich inputs into structured directives suitable for ML processing, (ii) enriching semantic representations to enhance decision-making quality, and (iii) dynamically optimizing learning pipelines and network configurations.
For example, \cite{zhang2024generativeai} introduced a Generative AI-in-the-Loop framework where LLMs convert intent-based network control commands into actionable tasks for ML models while simultaneously producing human-interpretable explanations. This work further demonstrated that feeding semantic knowledge generated by LLMs into ML-based intrusion detection systems (IDS) can significantly improve detection performance.
Similarly, \cite{cohen2025routing} introduced an algorithmic framework that integrates LLMs into large-scale ML training pipelines, allowing for network efficiency to be quantified. This framework periodically optimizes routing configurations to maintain global fairness and efficiency, ultimately achieving performance close to global optimality.
In the context of cybersecurity, \cite{hassanin2024llmcyber} demonstrated that combining LLMs with conventional ML-based IDS improves detection accuracy. This integration also enhances responsiveness to emerging threats and strengthens anomaly detection across diverse network environments.

\subsection{Hybrid LLM--Reinforcement Learning Approaches}
Hybrid LLM--reinforcement learning (LLM--RL) approaches aim to combine the agile decision-making capability of RL with reasoning, understanding, and feedback-generation ability of LLMs. LLM--RL integration focuses on improving the interaction loop among state perception, reward design, policy learning, and action refinement.

In \cite{du2024rlli}, LLM-empowered generative agents are used to simulate users with different personality traits and generate subjective QoE feedback, which serves as the reward signal for DRL-based artificial intelligence-generated content (AIGC) service-provider selection. This approach is useful when wireless service quality depends on subjective user satisfaction that is difficult to model analytically. In \cite{cai2026laeneet}, an LLM-enhanced RL framework is developed for low-altitude economy networking, where the LLM supports RL as an information processor, reward designer, decision-maker, and generator under energy, latency, coverage, and uncertainty constraints. The case study shows that the LLM-designed rewards can improve RL performance over manually designed rewards. For large-scale orchestration, \cite{shokrnezhad2025arc} proposes autonomous reinforcement coordination (ARC), which integrates LLMs with continual RL for semcom-enabled SAGIN resource orchestration. ARC uses an LLM-based high-level planner to sequence users and actions, while RL agents perform low-level resource-allocation decisions; RAG and replay-buffer management further support adaption to network dynamics.

Overall, these studies show three major LLM--RL integration patterns for wireless optimization: LLMs as reward or feedback generators, LLMs as semantic planners for RL agents, and LLMs as high-level coordinators in hierarchical RL systems. Such designs are promising for wireless scenarios where rewards are subjective, objectives are dynamic, and direct end-to-end optimization is difficult.

\subsection{Hybrid LLM--Evolutionary Approaches}
LLMs have recently become important components of evolutionary optimization frameworks. Unlike conventional evolutionary algorithms (EAs), which depend on handcrafted mutation, crossover, and selection operators with domain-specific tuning, LLMs enable semantic abstraction and zero-shot reasoning that can guide the search process through prompt-based mechanisms.

In the LMEA framework~\cite{liu2024lmea}, LLMs act as standalone optimizers by generating candidate solutions from prompts, using temperature modulation to balance exploration and exploitation. This approach achieves competitive results in combinatorial problems like the traveling salesman problem (TSP), without relying on population-based heuristics. For multi-objective optimization, the MOEA/D-LO model~\cite{liu2025large} integrates LLMs with decomposition-based search. Here, LLMs generate diverse offspring guided by subproblem-specific prompts. A white-box variant further distills interpretable optimization strategies from the LLM’s behavior, allowing insight into decision rules.
The AMO method~\cite{huang2025amo} advances toward autonomous optimization by chaining prompts and incorporating self-correction mechanisms. Feedback from the evaluation module informs new prompts in real time, allowing dynamic refinement of the optimization trajectory. Wu et al.~\cite{wu2025ecsurvey} provided a comprehensive survey, classifying recent developments into two categories: LLM-enhanced EAs and EA-enhanced LLMs. The former leverages LLMs for operator generation and adaptive control, while the latter applies EAs to fine-tune LLM architectures and outputs.

These hybrid frameworks offer new directions for interpretable and adaptable optimization, particularly under the complexity of wireless environments. However, several challenges remain, including high inference latency, difficulty in interpreting prompt influence, and the scalability of LLM-EA integration in resource-constrained edge systems.

\section{Conclusion}
This survey has explored the emerging role of LLMs in addressing optimization challenges within wireless communication. By organizing existing research into a functional taxonomy, we categorized LLM applications across core wireless resource management tasks such as power control, bandwidth allocation, mobile edge computing, beamforming, and ISAC. Additionally, we examined the prompt engineering strategies that enable effective LLM deployment. Our analysis of hybrid approaches integrating LLMs with RL, ML, and evolutionary computation points to a growing trend of combining strengths across paradigms to achieve greater system intelligence. Looking ahead, while LLMs offer significant promise, several challenges remain, such as model interpretability, real-time deployment, security, and the need for wireless-specific fine-tuning. Future research should focus on addressing these gaps while developing robust, scalable, and context-aware LLM frameworks tailored to the demands of wireless systems.

\bibliographystyle{unsrtnat}
\bibliography{IEEEabrv,ref2}

\end{document}